\documentclass[aps,prd,showpacs,superscriptaddress,amsmath,a4paper]{revtex4}
\usepackage{cancel}
\usepackage{graphicx}
\usepackage{color}
\usepackage{amsmath}
\usepackage{amssymb}

\newlength{\nseparation}
\newenvironment{nfigure}[1]
        {\begin{figure}[#1]\hrule\vspace{\nseparation}\par}
        {\vspace{\nseparation}\par \hrule \end{figure}}

\newcommand{\eq}[1]{Eq.~(\ref{#1})}

\begin{document}

\title{Effective Higgs Vertices in the generic MSSM}

\author{Andreas Crivellin}
\affiliation{Albert Einstein Center for Fundamental Physics, Institute for Theoretical Physics,\\
              University of Bern, CH-3012 Bern, Switzerland.}

\date{\today}

\begin{abstract}
In this article we consider chirally enhanced corrections to Higgs vertices in the most general MSSM. We include the contributions stemming from bilinear-terms, from the trilinear $A$-terms, and from their nonholomorphic analogues, the $A'$-terms, which couple squarks to the "wrong" Higgs field.
We perform a consistent renormalization of the Higgs vertices beyond the decoupling limit ($M_{\rm{SUSY}}\to\infty$), using a purely diagrammatic approach. The cancellation of the different contributions in and beyond the decoupling limit is discussed and the possible size of decoupling effects which occur if the SUSY particles are not much heavier than the electroweak-scale are examined.  In the decoupling limit we recover the results obtained in the effective-field-theory approach.
For the nonholomorphic $A'$-terms we find the well known $\tan\beta$ enhancement in the down-sector similar to the one for terms proportional to $\mu$. Due to the a priori generic flavor structure of these trilinear terms large flavor-changing neutral Higgs couplings can be induced. We also discover new $\tan\beta$ enhanced contributions involving the usual holomorphic $A$-terms, which were not discussed before in the literature. These corrections occur only if also flavor-diagonal nonholomorphic corrections to the Higgs couplings are present. This reflects the fact that the $A$-terms, and also the chirality-changing self-energies, are physical quantities and cannot be absorbed into renormalization constants. 
\end{abstract}

\pacs{11.10.Gh,12.15.Ff,12.60.Jv,14.80.Da}

\maketitle

\section{Introduction}

Chirally enhanced corrections to Higgs couplings in the MSSM have been under consideration for a long time starting with the first analysis of FCNC processes in Ref.~\cite{Hamzaoui:1998nu}. Due to the $\mu$-term in the MSSM superpotential, couplings to the "wrong" Higgs field are induced via quantum corrections. Therefore, in the decoupling limit in which all sparticles are heavy and integrated out, the MSSM is a two-Higgs-doublet (2HDM) model of type III. This means that even very heavy SUSY particles leave their imprint in the form of nonholomorphic Higgs-quark couplings. The resulting effective Higgs couplings are in general flavor-changing and are therefore of special interest since they can significantly enhance processes like $B_{s,d}\to \mu\mu$ \cite{Hamzaoui:1998nu,Babu:1999hn,Isidori:2001fv,Chankowski:2000ng,Buras:2002wq,Buras:2002vd,Dedes:2001fv,Bobeth:2002ch,Baek:2002rt,Mizukoshi:2002gs,Dedes:2002er,Carena:2006ai,Dedes:2008iw}. 
Especially the $\tan\beta$ enhanced corrections to Higgs couplings in the minimally-flavor violating MSSM have been under extensive investigation in the decoupling limit \cite{Hall:1993gn,Blazek:1995nv,Hamzaoui:1998nu,Babu:1999hn,Isidori:2001fv,Buras:2002wq,Buras:2002vd} and beyond \cite{Hofer:2009xb}. These couplings have also been examined in the general MSSM with a generic flavor structure using an effective-field-theory approach \cite{Isidori:2002qe}. 

\medskip

In this article we want to go beyond the analysis of Ref.~\cite{Isidori:2002qe} in several aspects. First we want to compute the renormalization of the Higgs couplings beyond the decoupling limit. For this purpose we use a purely diagrammatic approach with a tree-level definition of the super-CKM basis as we have applied before to the CKM matrix \cite{Crivellin:2008mq} and to the squark-quark-gluino vertex \cite{Crivellin:2009ar}. We also perform the calculation in the most general MSSM which in principle also contains, in addition to the holomorphic $A$-terms, the nonholomorphic $A'$-terms \cite{Hall:1990ac,Haber:2007dj}. By nonholomorphic we mean that these trilinear terms couple squarks to the "wrong" Higgs fields due to the following term in the soft-SUSY breaking Lagrangian \cite{Rosiek:1989rs,Rosiek:1995kg}:
\begin{equation}
L_{\rm{soft}}^{A'}  = h_u^{I*} \tilde q_{iL}^I {A'}_{ij}^d \tilde d_{jR}^*  + h_d^{I*} \tilde q_{iL}^I {A'}_{ij}^u \tilde u_{jR}^*  + h.c.\label{Aprime}
\end{equation}
For the calculation of the corrections to the Higgs vertices we will use a minimal renormalization scheme (MS or $\overline{\rm{MS}}$ for example). This has several advantages compared to an on-shell scheme as we want to illustrate later in more detail. First of all, the resummation of $\tan\beta$ turns out to be much easier in a minimal renormalization scheme. In addition, a minimal scheme corresponds to a tree-level definition of the super-CKM basis \cite{Crivellin:2008mq,Crivellin:2009ar} which allows for a direct relation between the parameters in the Lagrangian and physical quantities. 

\medskip

This article is structured as follows: First, in Sec. \ref{sec:Higgs} we quote the Feynman rules for the Higgs-quark and Higgs-squark vertices in our conventions. Section \ref{sec:Renormalization} is devoted to the diagrammatic calculation of the effective Higgs-quark vertices. In Sec. \ref{sec:EFT} we perform the analysis in the effective-field-theory confirming previous results in the decoupling limit. Here we also compare our results to the ones in the literature. Finally we conclude in Sec.~\ref{sec:conclusions}.

\section{Higgs Feynman Rules \label{sec:Higgs}}

\begin{nfigure}{t}
\includegraphics[width=0.8\textwidth]{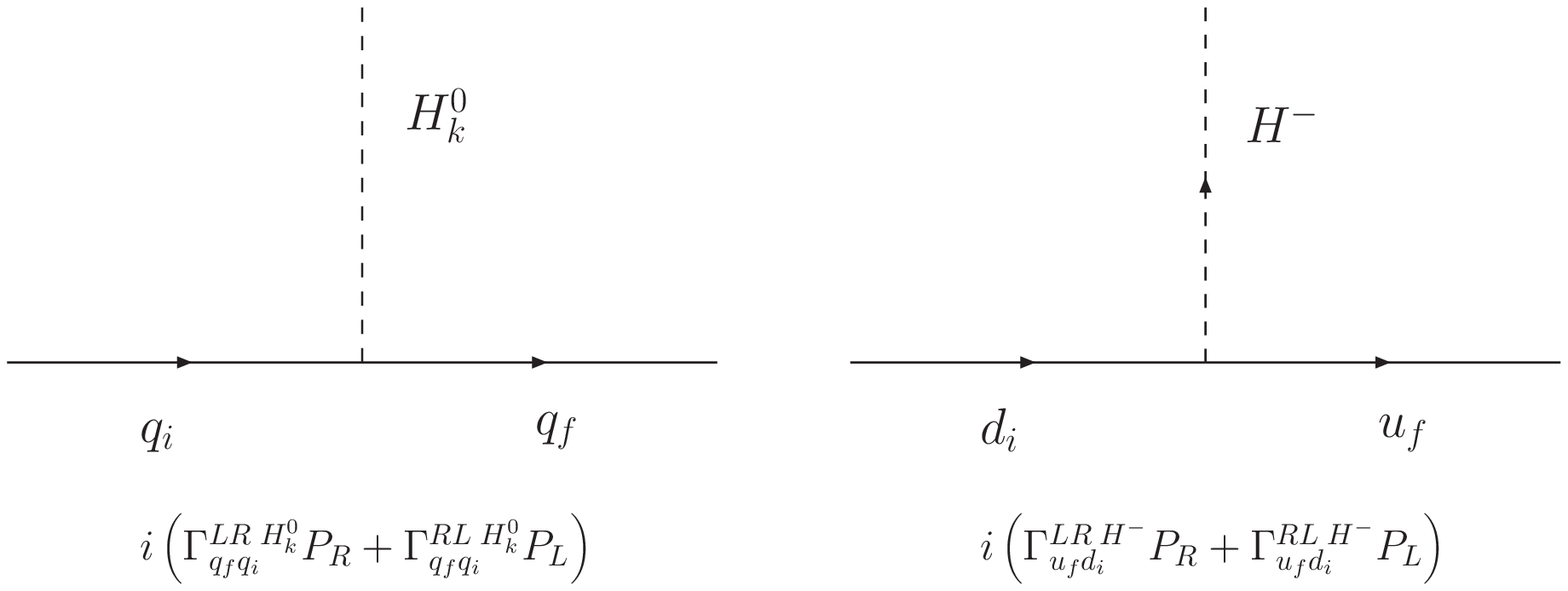}
\caption{Higgs-quark vertices with the corresponding Feynman-rules.}\label{fig:Higgs-Quark-Coupling}
\end{nfigure}

In this section we quote the Feynman-rules for the Higgs vertices both for easy reference for the reader and in order to fix our conventions and abbreviations. First consider the Higgs-quark vertices shown in Fig.~\ref{fig:Higgs-Quark-Coupling}. Let us write the Feynman-rules for the neutral Higgs as 
\begin{equation}
\renewcommand{\arraystretch}{1.8}
i\left( {\Gamma _{q_f q_i }^{RL\,H_k^0 } P_L  + \Gamma _{q_f q_i }^{LR\,H_k^0 } P_R } \right)\qquad\rm{with}\qquad H_k^0  = \left( {\begin{array}{*{20}c}
   {H^0 ,} & {h^0 ,} & {A^0 } . \\
\end{array}} \right)
\end{equation}
Here $H^0$ ($h^0$) is the heavy (light) CP-even Higgs and $A^0$ is the physical CP-odd Higgs particle. The indices $f$ and $i$ denote the flavors of the quarks. We also write the charged-Higgs vertex with an outgoing up-quark as
\begin{equation}
i\left( {\Gamma _{u_f q_i }^{RL\,H^ +  } P_L  + \Gamma _{u_f q_i }^{LR\,H^ +  } P_R } \right).
\end{equation}
With theses definitions we have the following couplings $\Gamma$ for the uncorrected (tree-level) vertices
\begin{equation}
\renewcommand{\arraystretch}{2.2}
\begin{array}{l}
 \Gamma _{q_f q_i }^{LR\,H_k^0 }  = \left( {\Gamma _{q_i q_f }^{RL\,H_k^0 } } \right)^*  = \tilde \Gamma _q^{H_k^0 } Y_i^q \delta _{fi}    ,
 \\ 
 \Gamma _{u_f d_i }^{LR\,H^ +  }  = \sin \left( \beta  \right)V_{fi}^{CKM\,\left( 0 \right)} Y_i^d , \\ 
 \Gamma _{u_f d_i }^{RL\,H^ +  }  = \cos \left( \beta  \right)Y_f^{u*} V_{fi}^{CKM\,\left( 0 \right)}.  \\ 
 \end{array}
 \label{GammaH}
\end{equation}
where we have defined the flavor independent quantity $\Gamma _q^{H_k^0 }$ which is given by:
\begin{equation}
\renewcommand{\arraystretch}{2.2}
\begin{array}{l}
 \tilde \Gamma _d^{H_k^0 }  = \left( {\dfrac{{ - 1}}{{\sqrt 2 }}\cos \left( \alpha  \right),\dfrac{1}{{\sqrt 2 }}\sin \left( \alpha  \right),\dfrac{i}{{\sqrt 2 }}\sin \left( \beta  \right)} \right), \\ 
 \tilde \Gamma _u^{H_k^0 }  = \left( {\dfrac{{ - 1}}{{\sqrt 2 }}\sin \left( \alpha  \right),\dfrac{{ - 1}}{{\sqrt 2 }}\cos \left( \alpha  \right),\dfrac{i}{{\sqrt 2 }}\cos \left( \beta  \right)} \right). \\ 
 \end{array}
\end{equation}
Here, $\alpha$ is the usual mixing angle of the Higgs sector (see for example \cite{Rosiek:1989rs,Rosiek:1995kg}) and $\tan\beta=v_u/v_d$ is the ratio of the vacuum expectation values acquired by $h_d$ and $h_u$, respectively. In \eq{GammaH}, $Y^q$ is the tree-level (uncorrected) Yukawa matrix of the MSSM superpotential which is diagonal in the super-CKM basis. We choose a "tree-level" definition of the super-CKM basis which respects Supersymmetry also at the loop-level \cite{Crivellin:2008mq,Crivellin:2009ar}. This means we diagonalize the bare Yukawa couplings and apply the same rotations to the squark fields. Therefore, the CKM matrix which occurs in the charged-Higgs vertex is the one which arises solely due to the misalignment between the bare Yukawa couplings. We will explain this in more detail in the next section.

\medskip

Now we consider the Higgs-squark couplings. Again we denote the Feynman-rule of the corresponding neutral Higgs vertex by $i\Gamma _{\tilde q_s \tilde q_t }^{H_k^0 }$ and the charged-Higgs vertex by $i\Gamma _{\tilde u_s \tilde d_t }^{H^ +  }$  where s (t) label the mass eigenstates of the outgoing (incoming) squark running from 1 to 6. In order to simplify the notation it is useful to define some more abbreviations:
\begin{align}
 \Gamma _{\tilde q_s \tilde q_t }^{H_k^0 \,EW}  &=  - a_q b_q^k \dfrac{{e^2 }}{{3c_W^2 }}\left( {\delta ^{st}  + \dfrac{{3 - 4a_q s_W^2 }}{{2a_q s_W^2 }}W_{is}^{\tilde d * } W_{it}^{\tilde d} } \right)\nonumber, \\ 
 \Gamma _{\tilde q_s \tilde q_t }^Y  &=  - v_q Y_q^i Y_q^{i * } \left( {W_{is}^{\tilde q * } W_{it}^{\tilde q}  + W_{i + 3,s}^{\tilde q * } W_{i + 3,t}^{\tilde q} } \right) \nonumber, \\ \nonumber
 \Gamma _{\tilde q_s \tilde q_t }^{RL\,A}  &=  - W_{j + 3,s}^{\tilde q * } A_{ij}^q W_{it}^{\tilde q},  \\ 
 \Gamma _{\tilde q_s \tilde q_t }^{LR\,A}  &=  - W_{is}^{\tilde q * } A_{ij}^{q * } W_{j + 3,t}^q \label{Gamma-Squark}, \\ \nonumber
 \Gamma _{\tilde q_s \tilde q_t }^{RL\,\mu A^\prime}  &=  - W_{j + 3,s}^{\tilde q * } \left( {A_{ij}^{\prime q * }  + \mu ^{*}  Y_i^{q *} \delta _{ij} } \right)W_{it}^{\tilde q} , \\ \nonumber
 \Gamma _{\tilde q_s \tilde q_t }^{LR\,\mu A^\prime}  &=  - W_{is}^{\tilde q * } \left( {A_{ij}^{\prime q}  + \mu Y_i^q \delta _{ij} } \right)W_{j + 3,t}^q . 
\end{align}
The definitions for the rotation matrices $W^{\tilde{q}}$ can be found in Ref.~\cite{Crivellin:2008mq} and the constants $a_q$ and $b^k_q$ are given by: 
\begin{equation}
\renewcommand{\arraystretch}{1.8}
\begin{array}{l}
 a_d  = 1,\,\;\;\;a_u  = 2,\;\;\;\; \\ 
 b_d^k  =  - b_u^k  = \left( {\begin{array}{*{20}c}
   {v_d \cos \left( \alpha  \right) - v_u \sin \left( \alpha  \right),} & { - v_d \sin \left( \alpha  \right) - v_u \cos \left( \alpha  \right),} & 0  \\
\end{array}} \right) .\\ 
 \end{array}
 \end{equation}
With these conventions the Feynman rules for the neutral Higgs vertices can be written in a very compact form:
\begin{equation}
\Gamma _{\tilde q_s \tilde q_t }^{H_k^0 }  = \Gamma _{\tilde q_s \tilde q_t }^{H_k^0 \,EW}  + \tilde \Gamma _q^{H_k^0 } \left( {\Gamma _{\tilde q_s \tilde q_t }^Y  + \Gamma _{\tilde q_s \tilde q_t }^{LR\,A}  + c_q^k \Gamma _{\tilde q_s \tilde q_t }^{LR\,\mu A'} } \right) + \tilde \Gamma _q^{H_k^{0 *}} \left( {\Gamma _{\tilde q_s \tilde q_t }^Y  + \Gamma _{\tilde q_s \tilde q_t }^{RL\,A}  + c_q^k \Gamma _{\tilde q_s \tilde q_t }^{RL\,\mu A'} } \right),
\label{Neutral-Higgs-Squark-vertex}
\end{equation}
with
\begin{equation}
c_d^k  = 1/c_u^k  = \left( {\tan \left( \alpha  \right), - \cot \left( \alpha  \right),-\cot \left( \beta  \right)} \right).
\end{equation}
Finally we have for the charged-Higgs:
\begin{equation}
\renewcommand{\arraystretch}{2.2}
 \begin{array}{*{20}c}
   \Gamma _{\tilde u_s \tilde d_t }^{H^ +  }  ={\left[ {\left( {\dfrac{{ - e^2 }}{{2s_W^2 }}\left( {v_d \sin \left( \beta  \right) + v_u \cos \left( \beta  \right)} \right) + v_d (Y_d^i )^2 \sin \left( \beta  \right) + v_u (Y_u^j )^2 \cos \left( \beta  \right)} \right)V_{ji}^{CKM\,(0)} W_{it}^{\tilde d} W_{js}^{\tilde u*} } \right.} \hfill  \\
 \phantom{\Gamma _{\tilde u_s \tilde d_t }^{H^ +  }  =}  {\,\;\; + \dfrac{{\sqrt 2 M_W s_W }}{e}Y_u^{j*} Y_d^i V_{ji}^{CKM\,(0)} W_{i + 3,t}^{\tilde d} W_{j + 3,s}^{\tilde u*} } \hfill  \\
 \phantom{\Gamma _{\tilde u_s \tilde d_t }^{H^ +  }  =}  {\,\;\; + \left( {\sin \left( \beta  \right)\mu ^* Y_u^{j*} \delta _{kj}  + \sin \left( \beta  \right)A_u^{\prime kj*}  - \cos \left( \beta  \right)A_u^{kj*} } \right)V_{ki}^{CKM\,(0)} W_{j + 3,s}^{\tilde u*} W_{it}^{\tilde d} } \hfill  \\
 \phantom{\Gamma _{\tilde u_s \tilde d_t }^{H^ +  }  =}  {\left. {\,\;\; + \left( { - \sin \left( \beta  \right)A_d^{ki}  + \cos \left( \beta  \right)A_d^{\prime ki}  + \cos \left( \beta  \right)\mu Y_d^i \delta _{ki} } \right)V_{jk}^{CKM\,(0)} W_{js}^{\tilde u*} W_{i + 3,t}^{\tilde d} } \right]} \hfill  \\
\end{array}
\end{equation}
With these vertices at hand, we can now calculate the effective Higgs-quark couplings including the chirally enhanced loop-corrections.

\section{Renormalization in the full theory\label{sec:Renormalization}}

We now want to calculate the effective Higgs-quark vertices taking into account the leading chirally enhanced corrections. We follow a diagrammatic approach treating all diagrams with flavor-changing self-energy on an external leg as one-particle irreducible \cite{Logan:2000iv}. (We already applied the same approach to the renormalization of the CKM matrix \cite{Crivellin:2008mq} and of the squark-quark-gluino vertex \cite{Crivellin:2009ar}.) These flavor-changing corrections can be viewed as rotation in flavor space, while flavor-conserving self-energies renormalize the quark masses.  In the following we will focus on the gluino contributions which are dominant (in most regions of parameter space) in the presence of generic sources of flavor-violation since they involve the strong coupling constant.

\medskip

As mentioned in the introduction we also include the nonholomorphic trilinear-terms which couple squarks to the "wrong" Higgs fields via the terms in \eq{Aprime}. These terms enter the squark mass matrices:
\begin{equation}
\renewcommand{\arraystretch}{2.6}
\begin{array}{l}
 {\rm M}_{\tilde u}^2  = \left( {\begin{array}{*{20}c}
   {V_{CKM}^{\left( 0 \right)} {\bf{M}}_{LL}^{\tilde q\,2} V_{CKM}^{\left( 0 \right)\dag }  + \dfrac{{\cos 2\beta }}{6}\left( {m_Z^2  + 2m_W^2 } \right)\hat 1 + v_u^2 Y^u Y^{u*} } & { - v_u {\bf{A}}^u  - v_d {\bf{A'}}^u  - v_u Y^u \mu \cot \beta }  \\
   { - v_u {\bf{A}}^{u\dag }  - v_d {\bf{A'}}^{u\dag }  - v_u Y^{u*} \mu ^* \cot \beta } & {{\bf{M}}_{RR}^{\tilde u\,2}  + \dfrac{{2\cos 2\beta }}{3}m_Z^2 \sin ^2 \theta _W {\bf{\hat 1}} + v_u^2 Y^u Y^{u*} }  \\
\end{array}} \right) \\ 
 {\rm M}_{\tilde d}^2  = \left( {\begin{array}{*{20}c}
   {{\bf{M}}_{LL}^{\tilde q\,2}  - \dfrac{{\cos 2\beta }}{6}\left( {m_Z^2  - 4m_W^2 } \right){\bf{\hat 1}} + v_d^2 Y^d Y^{d*} } & { - v_d {\bf{A}}_{}^d  - v_u {\bf{A'}}_{}^d  + v_d Y^d \mu \tan \beta }  \\
   { - v_d {\bf{A}}_{}^{d\dag }  - v_u {\bf{A'}}_{}^{d\dag }  - v_d Y^{d*} \mu ^* \tan \beta } & {{\bf{M}}_{RR}^{\tilde d\,2}  - \dfrac{{\cos 2\beta }}{3}m_Z^2 \sin ^2 \theta _W {\bf{\hat 1}} + v_d^2 Y^d Y^{d*} }  \\
\end{array}} \right) \\ 
 \end{array}
\end{equation}
Here $V^{(0)}_{\rm{CKM}}$ denotes the bare CKM matrix and $Y^q$ is the unrenormalized Yukawa coupling of the MSSM superpotential. Since the structure of the $A'$-terms in the squark mass matrices is similar to the term proportional to $\mu$ we can anticipate similar large $\tan\beta$ enhanced corrections for the couplings of down-quarks to Higgs particles.

\subsection{The SQCD quark self-energy}

\begin{nfigure}{t}
\includegraphics[width=0.5\textwidth]{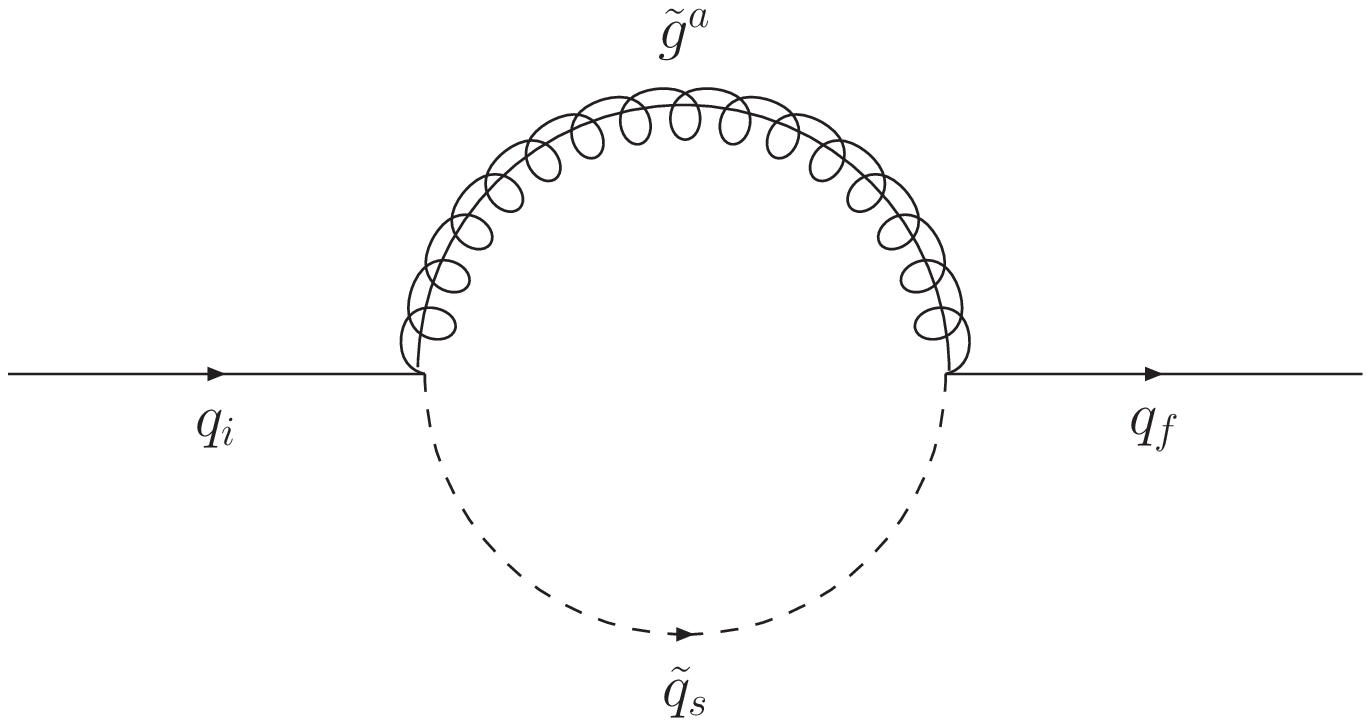}
\caption{Quark self-energy with squark and gluino as virtual particles. We receive the chirally enhanced part by evaluating this diagram at vanishing external momentum.}\label{fig:GluinoSelfEnergy}
\end{nfigure}

First, consider the quark self-energy with squarks and gluinos as virtual particles shown in Fig. \ref{fig:GluinoSelfEnergy}. In order to receive its chirally enhanced part it is sufficient to evaluate the diagram at vanishing external momentum \footnote{Terms which are of higher orders in the external momentum are suppressed by powers of $m_q^2/M_{\rm{SUSY}}^2$. For very light SUSY masses these terms could be relevant (non-decoupling) corrections in the case of an external top quark.}. With this simplification we get
\begin{equation}
\Sigma _{fi}^q  = \dfrac{2}{{3\pi }}\alpha _s m_{\tilde g} \sum\limits_{s,t = 1}^6 {\left( {V_{s\,fi}^{q\,LR} P_R  + V_{s\,fi}^{q\,RL} P_L } \right)B_0 \left( {m_{\tilde g}^2 ,m_{\tilde q_s }^2 } \right)} .
\end{equation}
Note that the self-energy is also non-decoupling. This means that it does not vanish for infinitely large SUSY masses. Since we know that the self-energy evaluated at vanishing external momentum is chirality-changing it must be proportional to at least one power of a chirality flipping off-diagonal $\Delta _{ij}^{q\,LR}$ element of the squark-mass matrix. Even though the mass-insertion approximation loses precision for extremely large off-diagonal elements in the squark mass matrix, the expansion in $\Delta_{ij}^{q\,AB}/M_{\rm{SUSY}}^2$ will always converge (at least slowly) since we know that the off-diagonal elements of the squark mass matrices are smaller than the diagonal ones (otherwise we would get negative squared squark masses). 

\medskip

In the decoupling limit the self-energy is linear in the chirality-flipping parameters. Therefore, it is possible (and useful) to factor out one power of $\Delta _{ij}^{q\,LR}$ and to write the self-energy in the following way:
\begin{equation}
\renewcommand{\arraystretch}{2.2}
\begin{array}{l}
 \Sigma _{fi}^{q\,LR}  = \dfrac{2}{{3\pi }}\alpha _s m_{\tilde g} \sum\limits_{j,k = 1}^3 {\sum\limits_{s,t = 1}^6 {V_{s\,fj}^{q\,LL} \Delta _{jk}^{q\,LR} V_{t\,ki}^{q\,RR} C_0 \left( {m_{\tilde g}^2 ,m_{\tilde q_s }^2 ,m_{\tilde q_t }^2 } \right)} },  \\ 
 \Sigma _{fi}^{q\,RL}  = \dfrac{2}{{3\pi }}\alpha _s m_{\tilde g} \sum\limits_{j,k = 1}^3 {\sum\limits_{s,t = 1}^6 {V_{s\,fj}^{q\,RR} \Delta _{kj}^{q\,LR*} V_{t\,ki}^{q\,LL} C_0 \left( {m_{\tilde g}^2 ,m_{\tilde q_s }^2 ,m_{\tilde q_t }^2 } \right)} }.  \\ 
 \end{array}
 \label{SQCD-SE}
\end{equation}
Note that this expression is exact in the decoupling limit. It contains all orders of chirality-conserving mass insertions since the rotation-matrices involved in $V_{s\,fj}^{q\,AB}$ take into account all possible flavor changes. One should keep in mind that in the decoupling limit, $V_{s\,fj}^{q\,AB}$ and $m_{\tilde q_s }$ do not depend on the chirality-flipping entries of the squark mass matrices. 

\medskip

Since 
\begin{equation}
\renewcommand{\arraystretch}{2.2}
\begin{array}{l}
 \Delta _{ij}^{d\,LR}  =  - v_d A_{ij}^d  - v_u A_{ij}^{\prime d}  - v_u \mu Y_i^d \delta _{ij},  \\ 
 \Delta _{ij}^{u\,LR}  =  - v_u A_{ij}^u  - v_d A_{ij}^{\prime u}  - v_d \mu Y_i^u \delta _{ij},  \\ 
 \end{array}
\end{equation}
we can split the self-energy into the following three parts:
\begin{equation}
\renewcommand{\arraystretch}{2.2}
\begin{array}{l}
 \Sigma _{fi\,A}^{q\,LR}  = \dfrac{{ - 2}}{{3\pi }}\alpha _s m_{\tilde g} v_d \sum\limits_{j,k = 1}^3 {\sum\limits_{s,t = 1}^6 {V_{s\,fj}^{q\,LL} A_{jk}^d V_{t\,ki}^{q\,RR} C_0 \left( {m_{\tilde g}^2 ,m_{\tilde q_s }^2 ,m_{\tilde q_t }^2 } \right)} } , \\ 
 \Sigma _{fi\,A'}^{q\,LR}  = \dfrac{{ - 2}}{{3\pi }}\alpha _s m_{\tilde g} v_u \sum\limits_{j,k = 1}^3 {\sum\limits_{s,t = 1}^6 {V_{s\,fj}^{q\,LL} A_{jk}^{\prime d} V_{t\,ki}^{q\,RR} C_0 \left( {m_{\tilde g}^2 ,m_{\tilde q_s }^2 ,m_{\tilde q_t }^2 } \right)} } , \\ 
 \Sigma _{fi\,Y}^{q\,LR}  = \dfrac{{ - 2}}{{3\pi }}\alpha _s m_{\tilde g} v_u \mu \sum\limits_{j = 1}^3 {\sum\limits_{s,t = 1}^6 {V_{s\,fj}^{q\,LL} Y_j^d V_{t\,ki}^{q\,RR} C_0 \left( {m_{\tilde g}^2 ,m_{\tilde q_s }^2 ,m_{\tilde q_t }^2 } \right)} }. \\ 
 \end{array}
 \label{eq:self-energy-decomposition}
\end{equation}
We derived these equations without having made an assumptions about the hierarchy between the SUSY and the electroweak scale (we just demand that the SUSY-particles are sufficiently heavier than the external quarks). In principle, the rotation matrices also depend on the $A$-terms and on the Yukawa coupling. Therefore, this decomposition is not unambiguous. However, in the decoupling limit no higher orders in $A$ or $Y^q$ survive and the rotation matrices depend only on the bilinear terms of the squark mass matrices. In this case it is useful to define the sum of all self-energies involving "wrong" Higgs couplings by:
\begin{equation}
\Sigma _{fi\,A^\prime \mu }^{q\,LR}  = \Sigma _{fi\,A^\prime}^{q\,LR}  + \Sigma _{fi\,Y}^{q\,LR} .
\end{equation}

\subsection{Minimal versus on-shell renormalization}

In this subsection we want to discuss and clarify the differences between minimal renormalization and on-shell renormalization regarding finite SUSY corrections. We emphasize that a minimal renormalization scheme leads to the same result for the bare quantities as the on-shell scheme, however, it is more straight forward and the symmetries of the superpotential are also manifest in the renormalized Yukawa couplings. 

\medskip

The self-energy in \eq{SQCD-SE} is finite. This means that the introduction of a counter-term is possible, but not necessary. There are two obvious choices: A minimal renormalization scheme (MS and $\overline{\rm{MS}}$ are equivalent here) or the on-shell scheme. Let us illustrate the two possibilities for the simple case of the renormalization of the bottom-quark mass by $\tan\beta$ enhanced corrections \footnote{In Ref.~\cite{Carena:1999py} it is also shown that one recovers the result for the $\tan\beta$ resummation of the effective-field-theory approach in the full theory using the on-shell scheme. We want to show in the following that one can obtain the same result in the full theory more directly using a minimal renormalization scheme. This shows that a minimal scheme allows for an easier and more direct comparison with the effective theory approach.}. In any renormalization scheme the physical quark mass $m_b$ is given by \footnote{Here we just consider the finite SUSY corrections}
\begin{equation}
	m_b=Y^b v_d+\delta Y^b v_d + Y^b v_u \varepsilon_b
	\label{mb_yb_deltab}
\end{equation}
$\delta Y^b$ being the counter-term to $Y^b$, $\varepsilon_b  = {\Sigma _{33\,Y_3^d }^{d\,LR} }/({Y_3^d v_u })$ and the bare Yukawa coupling is $Y^{b\,(0)}=Y^b +\delta Y^b $. The counter-term is determined by the renormalization condition. First consider the on-shell scheme (for details of the calculation see Ref.~\cite{Hofer:2009xb}). On-shell means that the physical bottom mass equals the renormalized one. This requirement determines the mass counter-term and the Yukawa counter-term at the one-loop level:
\begin{equation}
	\delta m_b\equiv v_d\delta Y^b=-Y^b v_u \varepsilon_b
	\label{deltaYb}
\end{equation}
Note that the counter-term can be of order one due to the $\tan\beta$ enhancement. At a higher loop-order $n$ there is only one chirally enhanced diagram which can also be of order one \cite{Carena:1999py}: the one-loop self-energy with the insertion of the $n-1$ counter term for $Y^b$. Therefore, we can iteratively solve for $Y^{b\,(0)}$, which is determined by a geometric series~\footnote{\eq{deltaYb} can also be solved directly in a minimal renormalization scheme by adding $m_b$ to both sides of the equation.}:
\begin{equation}
Y^{b\,(0)}=Y^{b}-\dfrac{m_b}{v_d}\left[\varepsilon_b\tan\beta-\left(\varepsilon_b\tan\beta\right)^2 +\left(\varepsilon_b\tan\beta\right)^3-...   \right]=\dfrac{m_b}{v_d\left(1+\varepsilon_b\tan\beta\right)}
\label{tanbetaresumm}
\end{equation}
This is the well know resummation formula for large $\tan\beta$ which determines the relation between the bare Yukawa coupling and the physical quark mass \footnote{The relation between the bottom Yukawa coupling and the bottom-quark mass has been computed including NNLO corrections \cite{Guasch:2003cv,Noth:2010jy}.}.

\medskip

Now we want to derive the corresponding formula in a minimal renormalization scheme where things will turn out to be much simpler. Minimal renormalization implies that we don't introduce a counter-term at all since the corrections are finite. Therefore, the physical bottom-quark mass is just given by
\begin{equation}
m_{b }  = v_d Y^b  + Y^b v_u \varepsilon_b.
\label{Yb-mb-MS}
\end{equation}
Since the counter-term is put to zero, the renormalized mass (Yukawa coupling) is equal to the bare mass (Yukawa coupling). However, the renormalized quark mass is no longer the physical one but rather equals the part of the mass which originates at tree-level from the Yukawa coupling of the superpotential. This statement is valid to all orders in perturbation theory. Since we don't introduce a counter-term, there is also nothing we could have to insert at higher loop-orders. Even though the meaning of the renormalized quantities is different now, the relation between the physical quark mass for the bare Yukawa coupling remains unchanged. This is easy to see, we just have to solve \eq{Yb-mb-MS} for the Yukawa in order to recover \eq{tanbetaresumm}. This simple example illustrates that a minimal renormalization scheme is simpler for the conceptional point of view since no higher loop-diagrams have to be taken into account, and therefore the resummation is automatically achieved. 

\medskip

The same arguments also apply to the flavor-changing case. One can, in principle, cancel flavor-changing self-energies on external lines with flavor-off-diagonal mass counter-terms. However, these counter-terms reappear in the vertices via the LSZ factor. In the case of the $W$-vertex these counter-terms, or equivalently the self-energy correction can be absorbed into the CKM matrix \cite{Crivellin:2008mq}. Again an on-shell or a minimal renormalization is possible. In this case the same arguments apply as in the case of the bottom-quark mass. Also in the case of the renormalization of the squark-quark-gluino vertex a minimal renormalization scheme is preferred. We have already discussed this in detail in Ref.~\cite{Crivellin:2009ar}. In addition, in the limit when masses or CKM mixing angles are generated radiatively \cite{Crivellin:2010gw} it is unnatural to introduce counter-terms to quantities which are zero at tree-level because in this case the counter-terms would break the symmetry of the Lagrangian. Of course, minimal renormalization and on-shell renormalization lead to the same physical results, however, due to the arguments presented above we will use a minimal renormalization scheme in the following which we believe is clearly preferred over an on-shell one.

\subsection{Finite renormalization of quark masses and wave functions}

Now let us consider the general finite renormalization of wave functions and quark masses in a minimal renormalization scheme. The physical quark mass is given by
\begin{equation}
m_{q_i }  = v_q Y_i^q  + \Sigma _{ii}^{q\,LR} \label{mq-Yq}.
\end{equation}
Equation~\ref{mq-Yq} determines implicitly the Yukawa couplings for given SUSY parameters. The self-energy on the right side can in principle contain arbitrarily many powers of Yukawa couplings. In this case it is only possible to determine the Yukawa coupling analytically in the absence of flavor-mixing \cite{Hofer:2009xb}. However, \eq{mq-Yq} can be easily solved numerically since, as already noted, the mass-insertion approximation has to converge due to the positivity condition for the squark masses which demands that the off-diagonal elements are smaller than the diagonal ones. However, it is useful to have an analytic formula at hand. If we restrict ourselves to terms proportional to only one power of $Y^d$ (which is exact in the decoupling limit) than we recover the well known resummation formula for $\tan\beta$ enhanced corrections in the down-quark sector with a correction due to the $A$-terms
\begin{equation}
Y_i^d  = \dfrac{{m_{d_i }  - \Sigma _{ii\,\cancel{Y_i^d}}^{d\,LR} }}{{v_d \left( {1 + \tan \left( \beta  \right)\varepsilon _i^d } \right)}}.
\label{md-Yd}
\end{equation}
Here $\varepsilon _i^d$ is given by 
\begin{equation}
\varepsilon _i^d  = \dfrac{{\Sigma _{ii\,Y_i^d }^{d\,LR} }}{{Y_i^d v_u }} .
\label{eq:epsilon_d}
\end{equation}
and $\Sigma _{ii\,\cancel{Y_i^d}}^{d\,LR}$ is the part of the self-energy which involves no Yukawa coupling $Y^d_i$ (for example also $\Sigma _{11}^{d\,LR}  \sim \Delta _{13}^{LL} v_d Y_3^d \tan \left( \beta  \right)\Delta _{31}^{RR} $ is included here).
In the up-sector we can safely neglect the self-energy contributions proportional to $Y^u$ since they are suppressed by $\cot(\beta)$. Therefore, the Yukawa coupling is simply given by
\begin{equation}
Y_i^u  = \left(m_{u_i }  - \Sigma _{ii}^{u\,LR}\right)/v_u .
\label{mu-Yu}
\end{equation}
Furthermore, the flavor-changing self-energies induce also a wave-function rotation in flavor space. This rotation has to be applied to all external quark fields. At the one loop level, neglecting small mass-ratios, it is given by \cite{Crivellin:2008mq}:
\begin{equation}
\renewcommand{\arraystretch}{2.2}
U_{fi}^{q\,L}  = \left( {\begin{array}{*{20}c}
   1 & {\dfrac{{\Sigma _{12}^{q\,LR} }}{{m_{q_2 } }}} & {\dfrac{{\Sigma _{13}^{q\,LR} }}{{m_{q_3 } }}}  \\
   {\dfrac{{ - \Sigma _{21}^{q\,RL} }}{{m_{q_2 } }}} & 1 & {\dfrac{{\Sigma _{23}^{q\,LR} }}{{m_{q_3 } }}}  \\
   {\dfrac{{ - \Sigma _{31}^{q\,RL} }}{{m_{q_3 } }}} & {\dfrac{{ - \Sigma _{32}^{q\,RL} }}{{m_{q_3 } }}} & 1  \\
\end{array}} \right)_{fi}
\label{DeltaU}
\end{equation}
However, for transitions between the third and the first generation also two-loop corrections can be important \cite{Crivellin:2008mq,Crivellin:2010gw}. Applying the rotations in \eq{DeltaU} to the W-vertex renormalizes the CKM matrix. Then bare CKM matrix (stemming from the misalignment between the Yukawa couplings) can now be calculated in terms of the physical one:
\begin{equation}
V^{CKM\left( 0 \right)}  = U^{u\,L\dag } V^{CKM} U^{d\,L} 
\label{CKM-0-ren}
\end{equation}

\medskip

Here a comment on the quark mass appearing propagator is in order. Without the self-energy corrections, the propagator contains the bare quark mass $Y^q_i v_q$. However, the self-energy corrections have to be included to all orders using the Dyson resummation. In this way, again the physical mass $M_{q_i}=Y^q_i v_q+\Sigma _{ii}^{q\,LR}$ appears in the propagator. This mass, which enters in \eq{DeltaU} has to be evaluated at the same scale as the self-energy corrections $\Sigma^{q\,LR}_{ij}$. Furthermore, it can be shown that it is the $\overline{\rm{MS}}$ renormalized quark mass \cite{Hofer:2009xb} extracted from experiment using the SM prescription. 

\subsection{Calculation of the effective Higgs-vertices}

Now we are ready to address the renormalization of the Higgs-quark vertices. First, let us apply the field rotation in \eq{DeltaU} to the neutral Higgs vertices. If we do this, the self-energy contributions to the neutral and charged-Higgs vertices are simply given by: 
\begin{equation}
\renewcommand{\arraystretch}{2.2}
\begin{array}{*{20}c}
   {\left. {\Gamma _{q_f q_i }^{RL\,H_k^0 } } \right|_{SE}^{eff}  = U_{jf}^{q\,R*} \Gamma _{q_j q_k }^{RL\,H_k^0 } U_{ki}^{q\,L} }, \qquad & {\left. {\Gamma _{q_f q_i }^{LR\,H_k^0 } } \right|_{SE}^{eff}  = U_{jf}^{q\,L*} \Gamma _{q_j q_k }^{LR\,H_k^0 } U_{ki}^{q\,R} } , \\
   {\left. {\Gamma _{u_f d_i }^{RL\,H^ +  } } \right|_{SE}^{eff}  = U_{jf}^{u\,R*} \Gamma _{u_j d_k }^{RL\,H^ +  } U_{ki}^{d\,L} }, \qquad & {\left. {\Gamma _{u_f d_i }^{LR\,H^ +  } } \right|_{SE}^{eff}  = U_{jf}^{u\,L*} \Gamma _{u_j d_k }^{LR\,H^ +  } U_{ki}^{d\,R} } . \\
\end{array}
\end{equation}
This means that in order to obtain the flavor structure of the self-energy corrections we have to calculate:
\begin{equation}
\renewcommand{\arraystretch}{2.4}
\tilde Y_{fi}^q  = U_{jf}^{q\,L*} Y_j^q U_{ji}^{q\,R}  = Y^{q_i } \delta _{fi}  - \left( {\begin{array}{*{20}c}
   0 & {\dfrac{{Y_2^q }}{{m_{q_2 } }}\Sigma _{12}^{q\,LR} } & {\dfrac{{Y_3^q }}{{m_{q_2 } }}\Sigma _{13}^{q\,LR} }  \\
   {\dfrac{{Y_2^q }}{{m_{q_2 } }}\Sigma _{21}^{q\,LR} } & 0 & {\dfrac{{Y_3^q }}{{m_{q_2 } }}\Sigma _{23}^{q\,LR} }  \\
   {\dfrac{{Y_3^q }}{{m_{q_2 } }}\Sigma _{31}^{q\,LR} } & {\dfrac{{Y_3^q }}{{m_{q_2 } }}\Sigma _{32}^{q\,LR} } & 0  \\
\end{array}} \right)_{fi}
\label{Ytilde}
\end{equation}
Substituting \eq{md-Yd} and \eq{mu-Yu} into \eq{Ytilde} we can express everything via self-energies and physical masses:
\begin{eqnarray}
\renewcommand{\arraystretch}{2.6}
 \tilde Y_{fi}^u  = \dfrac{1}{{v_u }}\left( {m_{u_i } \delta _{fi}  - \Sigma _{fi}^{u\,LR}  + \left( {\begin{array}{*{20}c}
   0 & {\dfrac{{\Sigma _{22}^{u\,LR} }}{{m_{u_2 } }}\Sigma _{12}^{u\,LR} } & {\dfrac{{\Sigma _{33}^{u\,LR} }}{{m_{u_3 } }}\Sigma _{13}^{u\,LR} }  \\
   {\dfrac{{\Sigma _{22}^{u\,LR} }}{{m_{u_2 } }}\Sigma _{21}^{u\,LR} } & 0 & {\dfrac{{\Sigma _{33}^{u\,LR} }}{{m_{u_3 } }}\Sigma _{23}^{u\,LR} }  \\
   {\dfrac{{\Sigma _{33}^{u\,LR} }}{{m_{u_3 } }}\Sigma _{31}^{u\,LR} } & {\dfrac{{\Sigma _{33}^{u\,LR} }}{{m_{u_3 } }}\Sigma _{32}^{u\,LR} } & 0  \\
\end{array}} \right)_{fi}} \right)  \\ 
 \tilde Y_{fi}^d  = \dfrac{1}{{v_d }}\dfrac{1}{{1 + \tan \left( \beta  \right)\varepsilon _{\max \left( {f,i} \right)}^d }}\left( {\begin{array}{*{20}c}
   {m_{d_1 }  - \Sigma _{11\,\cancel{Y^d_1}}^{d\,LR} } & {\dfrac{{m_{d_2 }  - \Sigma _{22\,\cancel{Y^d_2}}^{d\,LR} }}{{m_{d_2 } }}\Sigma _{12}^{d\,LR} } & {\dfrac{{m_{d_3 }  - \Sigma _{33\,\cancel{Y^d_3}}^{d\,LR} }}{{m_{d_3 } }}\Sigma _{13}^{d\,LR} }  \\
   {\dfrac{{m_{d_2 }  - \Sigma _{22\,\cancel{Y^d_2}}^{d\,LR} }}{{m_{d_2 } }}\Sigma _{21}^{d\,LR} } & {m_{d_2 }  - \Sigma _{22\,\cancel{Y^d_2}}^{d\,LR} } & {\dfrac{{m_{d_3 }  - \Sigma _{33\,\cancel{Y^d_3}}^{d\,LR} }}{{m_{d_3 } }}\Sigma _{23}^{d\,LR} }  \\
   {\dfrac{{m_{d_3 }  - \Sigma _{33\,\cancel{Y^d_3}}^{d\,LR} }}{{m_{d_3 } }}\Sigma _{31}^{d\,LR} } & {\dfrac{{m_{d_3 }  - \Sigma _{33\,\cancel{Y^d_3}}^{d\,LR} }}{{m_{d_3 } }}\Sigma _{32}^{d\,LR} } & {m_{d_3 }  - \Sigma _{33\,\cancel{Y^d_3}}^{d\,LR} }  \\
\end{array}} \right)_{fi} 
\end{eqnarray}
Here we have neglected small mass ratios and terms which involve more than one flavor-changing self-energy. Using \eq{CKM-0-ren} we can derive the following equalities which appear in charged-Higgs vertices
\begin{equation}
\renewcommand{\arraystretch}{2.2}
\begin{array}{l}
 U_{}^{u\,L\dag } V^{CKM\,\left( 0 \right)} Y_{}^{d\,\left( 0 \right)} U^{d\,R}  = V^{CKM} U_{}^{d\,L\dag } Y^{d\,\left( 0 \right)} U_{}^{d\,R},  \\ 
 U_{}^{u\,R\dag } Y_{}^{u*\,\left( 0 \right)} V^{CKM\,\left( 0 \right)} U^{d\,L}  = U_{}^{u\,R\dag } Y^{u\,\left( 0 \right)*} U_{}^{u\,L} V^{CKM}.  \\ 
 \end{array}
\end{equation}
In this way the quantities $\tilde{Y}^q$ also occur in the charged-Higgs vertices. Hence, we find for the self-energy corrections to the Higgs vertices
\begin{equation}
\left. {\Gamma _{q_f q_i }^{LR\,H_k^0 } } \right|_{SE}^{eff}  = U_{jf}^{q\,L*} \Gamma _{q_j }^{LR\,H_k^0 } U_{ji}^{q\,R}  = \tilde Y_{fi}^q \tilde \Gamma _q^{H_k^0 } ,
\label{Neutral-Higgs-SE}
\end{equation}
\begin{equation}
\renewcommand{\arraystretch}{2.2}
\begin{array}{l}
 \left. {\Gamma _{u_f d_i }^{LR\,H^ +  } } \right|_{SE}^{eff}  = \sin \left( \beta  \right)V_{fj}^{CKM} \tilde Y_{ji}^d,  \\ 
 \left. {\Gamma _{u_f d_i }^{RL\,H^ +  } } \right|_{SE}^{eff}  = \cos \left( \beta  \right)\tilde Y_{jf}^{u*} V_{ji}^{CKM}.  \\ 
 \end{array}
 \label{Charged-Higgs-SE}
\end{equation}
This completes the calculation of the self-energy corrections. Next we turn to the genuine vertex corrections. As explained in the previous subsection, the rotations in flavor-space $U_{ij}^{q\,L,R}$  can be of order one in the presence of chirally-enhanced flavor changing self-energies. The same is true concerning the genuine vertex correction. Therefore, a simple power-counting of $\alpha_s$ is not possible since terms proportional to $\alpha_s^2$ can be of the same order as terms proportional to $\alpha_s$. This means, we have to apply the rotation matrices $U^{q\,L,R}_{ij}$ to all external quarks, also the ones in the genuine vertex correction. This is crucial in order to obtain the correct vertices in the decoupling limit. As it turns out, without these additional rotations, we would not reproduce the Higgs couplings obtained in the effective-field-theory approach (see Sec.~\ref{sec:EFT}). Hence the genuine vertex corrections including theses rotations read for the neutral Higgs couplings
\begin{equation}
\renewcommand{\arraystretch}{2.2}
\begin{array}{l}
 \left. {\Gamma _{q_f q_i }^{LR\,H_k^0 } } \right|_{GVren}^{eff}  = \dfrac{2}{{3\pi }}\alpha _s m_{\tilde g} \sum\limits_{s,t = 1}^6 {\sum\limits_{j,l = 1}^3 {U_{jf}^{q\,L*} W_{js}^{\tilde q} \Gamma _{st}^{H_k^0 } W_{l + 3,t}^{\tilde q*} U_{li}^{q\,R} C_0 \left( {m_{\tilde g}^2 ,m_{\tilde q_t }^2 ,m_{\tilde q_s }^2 } \right)} },  \\ 
 \left. {\Gamma _{q_f q_i }^{RL\,H_k^0 } } \right|_{GVren}^{eff}  = \dfrac{2}{{3\pi }}\alpha _s m_{\tilde g} \sum\limits_{s,t = 1}^6 {\sum\limits_{j,l = 1}^3 {U_{jf}^{q\,R*} W_{j + 3,s}^{\tilde q} \Gamma _{st}^{H_k^0 } W_{lt}^{\tilde q*}  U_{li}^{q\,L} C_0 \left( {m_{\tilde g}^2 ,m_{\tilde q_t }^2 ,m_{\tilde q_s }^2 } \right)} } , \\ 
 \end{array}
  \label{Neutral-Higgs-GV}
\end{equation}
and for the charged-Higgs
\begin{equation}
\renewcommand{\arraystretch}{2.2}
\begin{array}{l}
 \left. {\Gamma _{u_f d_i }^{LR\,H^ +  } } \right|_{GVren}^{eff}  = \dfrac{2}{{3\pi }}\alpha _s m_{\tilde g} \sum\limits_{s,t = 1}^6 {\sum\limits_{j,l = 1}^3 {U_{jf}^{u\,L*} W_{js}^{\tilde u} \Gamma _{st}^{H^ +  } W_{l + 3,t}^{\tilde d*} U_{li}^{d\,R} C_0 \left( {m_{\tilde g}^2 ,m_{\tilde u_s }^2 ,m_{\tilde d_t }^2 } \right)} },  \\ 
 \left. {\Gamma _{u_f d_i }^{RL\,H^ +  } } \right|_{GVren}^{eff}  = \dfrac{2}{{3\pi }}\alpha _s m_{\tilde g} \sum\limits_{s,t = 1}^6 {\sum\limits_{j,l = 1}^3 {U_{jf}^{u\,R*} W_{j + 3,s}^{\tilde u} \Gamma _{st}^{H^ +  } W_{lt}^{\tilde d*} U_{li}^{d\,L} C_0 \left( {m_{\tilde g}^2 ,m_{\tilde u_s }^2 ,m_{\tilde d_t }^2 } \right)} } . \\ 
 \end{array}
 \label{Charged-Higgs-GV}
\end{equation}
We obtain the complete effective neutral (charged) Higgs couplings by adding \eq{Neutral-Higgs-GV} to \eq{Neutral-Higgs-SE} (\eq{Charged-Higgs-GV} to \eq{Charged-Higgs-SE}).

\subsection{Higgs-vertices in the decoupling limit}

Now we calculate the effective Higgs vertices in the decoupling limit. In this limit all surviving terms must involve nonholomorphic corrections. This means these must terms involve "wrong" Higgs couplings (i.e. go beyond a type-II 2HDM). We will verify the result in the next section using an effective-field-theory approach. In the decoupling limit the genuine vertex corrections simplify to:
\begin{equation}
\renewcommand{\arraystretch}{2.2}
\begin{array}{l}
 \left. {\Gamma _{d_f d_i }^{LR\,H_k^0 } } \right|_{GV}  = \tilde \Gamma _d^{H_k^0 } \left( {\dfrac{{\tilde \Sigma _{fi\,A}^{d\,LR} }}{{v_d }} + c_d^k \dfrac{{\tilde \Sigma _{fi\,A^{\prime}\mu }^{d\,LR} }}{{v_u }}} \right), \\ 
 \left. {\Gamma _{u_f u_i }^{LR\,H_k^0 } } \right|_{GV}  = \tilde \Gamma _u^{H_k^0 } \left( {\dfrac{{\tilde \Sigma _{fi\,A}^{u\,LR} }}{{v_u }} + c_u^k \dfrac{{\tilde \Sigma _{fi\,A^{\prime}\mu }^{u\,LR} }}{{v_d }}} \right), \\ 
 \left. {\Gamma _{u_f d_i }^{LR\,H^ \pm  } } \right|_{GV}^{eff}  = \dfrac{1}{v}\sum\limits_{j = 1}^3 {V_{fj}^{CKM\,\left( 0 \right)} \left( {\tan \left( \beta  \right)\Sigma _{ji\,A}^{d\,LR}  + \cot \left( \beta  \right)\Sigma _{ji\,A^\prime \mu}^{d\,LR}   } \right)},  \\ 
 \left. {\Gamma _{u_f d_i }^{RL\,H^ \pm  } } \right|_{GV}^{eff}  = \dfrac{1}{v}\sum\limits_{j = 1}^3{\left( {\cot \left( \beta  \right)\Sigma _{fj\,A}^{u\,RL}  + \tan \left( \beta  \right)\Sigma _{fj\,A^\prime \mu}^{u\,RL}  } \right)} V_{ji}^{CKM\,\left( 0 \right)}.  \\ 
 \end{array}
 \label{GV-decoupling}
\end{equation}
Here, we have expressed the charged-Higgs coupling in terms of the physical CKM matrix and used the fact that in the decoupling limit, the left-left blocks of the up and down squark mass matrices just differ by a CKM rotation, which implies:
\begin{equation}
\renewcommand{\arraystretch}{2.2}
	\begin{array}{c}
 V_{s\,fj}^{u\,LL} V_{jk}^{CKM(0)}  = V_{fj}^{CKM(0)} V_{s\,jk}^{d\,LL},  \\ 
 V_{kl}^{CKM(0)} V_{t\,li}^{d\,LL}  = V_{t\,kl}^{u\,LL} V_{li}^{CKM(0)}.  \\ 
 \end{array}
\end{equation}
In \eq{GV-decoupling} the quantity $\tilde \Sigma _{fi\,X}^{q\,LR}$ (with $X=A^\prime$ or $X=\mu$) contains the effects of external flavor-changing self-energies and is given by 
\begin{equation}
\renewcommand{\arraystretch}{2.2}
U_{jf}^{q\,L*} \Sigma _{jk\,X}^{q\,LR} U_{ki}^{q\,R}  = \tilde \Sigma _{jk\,X}^{q\,LR}  = \Sigma _{fi\,X}^{q\,LR}  - \left( {\begin{array}{*{20}c}
   0 & {\dfrac{{\Sigma _{22\,X}^{q\,LR} }}{{m_{q_2 } }}\Sigma _{12}^{u\,LR} } & {\dfrac{{\Sigma _{33\,X}^{q\,LR} }}{{m_{q_3 } }}\Sigma _{13}^{q\,LR} }  \\
   {\dfrac{{\Sigma _{22\,X}^{q\,LR} }}{{m_{q_2 } }}\Sigma _{21}^{u\,LR} } & 0 & {\dfrac{{\Sigma _{33\,X}^{q\,LR} }}{{m_{q_3 } }}\Sigma _{23}^{q\,LR} }  \\
   {\dfrac{{\Sigma _{33\,X}^{q\,LR} }}{{m_{q_3 } }}\Sigma _{31}^{q\,LR} } & {\dfrac{{\Sigma _{33\,X}^{q\,LR} }}{{m_{q_3 } }}\Sigma _{32}^{q\,LR} } & 0  \\
\end{array}} \right)_{fi}.
\label{Sigma-tilde}
\end{equation}
We can now add the genuine vertex correction to the self-energy contributions and find
\begin{equation}
\renewcommand{\arraystretch}{2.6}
\begin{array}{c}
 \left. {\Gamma _{u_f u_i }^{LR\,H_k^0 } } \right|_{dec}^{eff}  = \tilde \Gamma _u^{H_k^0 } \dfrac{1}{{v_u }}\left( {m_{u_i } \delta _{fi}  - \tilde \Sigma _{fi\,A^\prime\mu}^{u\,LR} \left( {1 - \tan \left( \beta  \right)c_u^k } \right)} \right) \\ 
 \left. {\Gamma _{d_f d_i }^{LR\,H_k^0 } } \right|_{dec}^{eff}  = \tilde \Gamma _d^{H_k^0 } \dfrac{1}{{v_d }}\left( {m_{d_i } \delta _{fi}  - \tilde\Sigma _{fi\,A^\prime\mu }^{d\,LR} \left( {1 - \cot \left( \beta  \right)c_d^k } \right)} \right) \\ 
 \left. {\Gamma _{u_f d_i }^{LR\,H^ \pm  } } \right|_{dec}^{eff}  = \dfrac{1}{v}\sum\limits_{j = 1}^3 {V_{fj}^{CKM} \left( {-\left( {\cot \left( \beta  \right) + \tan \left( \beta  \right)} \right)\tilde \Sigma _{ji\,A^\prime\mu}^{ d\,LR}  + \tan \left( \beta  \right)m_{d_i } \delta _{ji} } \right)}  \\ 
 \left. {\Gamma _{u_f d_i }^{RL\,H^ \pm  } } \right|_{dec}^{eff}  = \dfrac{1}{v}\sum\limits_{j = 1}^3 {\left( {-\left( {\tan \left( \beta  \right) + \cot \left( \beta  \right)} \right)\tilde \Sigma _{fj\,A^\prime \mu}^{u\,RL} + \cot \left( \beta  \right)m_{u_f } \delta _{fj} } \right)} V_{ji}^{CKM}.  \\ 
 \end{array}
 \label{Higgs-vertices-decoupling}
\end{equation}
The coupling to down-type quarks depends implicitly on the Yukawa coupling. We can express everything in terms of SUSY parameters and physical masses using \eq{md-Yd} which implies:
\begin{equation}
\Sigma _{ij\,A^\prime \mu }^{ d\,LR}  = \Sigma _{ij\,A^\prime}^{ d\,LR}  + \frac{{\left( {m_{d_i }  - \Sigma _{ii\,A}^{d\,LR} } \right)\varepsilon _i^d \tan \left( \beta  \right)}}{{1 + \tan \left( \beta  \right)\varepsilon _i^d }}\delta _{ij} 
\label{md-Yd2}
\end{equation}
Inserting \eq{md-Yd2} into \eq{Higgs-vertices-decoupling} we recover the well known $\tan\beta$ enhanced correction to the bottom-quark mass in the absence of flavor-violation. 
In the up-sector we can safely neglect terms proportional to $\mu$ since they are also proportional to quark masses and $\cot\beta$. However, although the $A^{\prime u}$-terms also come with $v_d$ they can still be relevant due to their generic flavor structure. Note that the cancellation between the self-energy contributions and the genuine vertex diagram in the case $\mu=A^\prime=0$ observed in this section, is related to the fact that one must get a type-II 2HDM in the decoupling limit. This means that in the absence of nonholomorphic corrections the effect of the $A$-terms can be absorbed into a effective Yukawa coupling \footnote{This cancellation has been already observed for the flavor-conserving case considering $A^b$ in Ref.~\cite{Guasch:2003cv}.}. 

\medskip

Note that in the couplings to down quarks in \eq{Higgs-vertices-decoupling} there is a new contribution not discovered before in the literature due to the last term in \eq{Sigma-tilde} which is a combination of a flavor-diagonal nonholomorphic term with a flavor-changing one. Diagrammatically, this term remains because of an imperfect cancellation between the Yukawa coupling in Higgs-quark vertex and the quark mass in the denominator of the propagator, since at large $\tan\beta$ the ratio $Y^d_i/(m_{d_i}/v_d)$ is unequal to one (see Fig.~\ref{fig:eff_higgs_vertex}). This new contribution is numerically important for large values of $\tan(\beta)$ since the flavor structure of the A-terms directly enters the self-energies, which does not need to involve further nonholomorphic terms. 

\begin{nfigure}{t}
\includegraphics[width=0.8\textwidth]{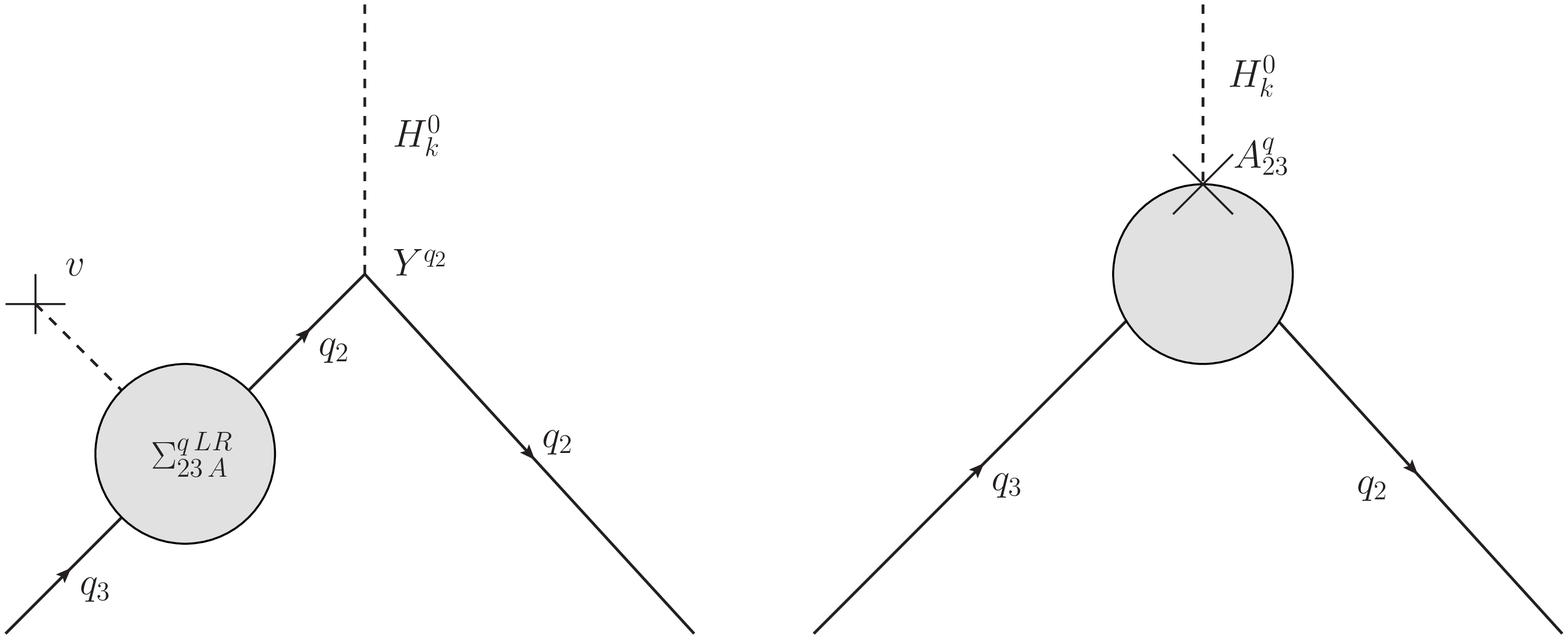}
\caption{Self-energy and genuine vertex correction involving
  $A^q_{23}$ contributing to the effective Higgs
  coupling. The cancellation between the two diagrams in imperfect since $Y^q_i v_q\neq m_{q_i}$.}\label{fig:eff_higgs_vertex}
\end{nfigure}

\subsection{Decoupling Corrections\label{sec:decoupling-corrections}}

Finally, we discuss the cancellations between the self-energy contributions and the genuine vertex correction beyond the decoupling limit and quantify the possible size of the decoupling effects. We will do this for the neutral Higgs vertices. The generalization to the charged-Higgs couplings is straight forward and no new effects occur. 

\medskip

There are two different types of decoupling effects. First, there are terms which are leftover, if one adds the genuine vertex correction to the self-energy contributions beyond the decoupling limit. These terms correspond to higher dimensional operators and do not match on the SU(2)-invariant structure of the 2HDM. Second, there are decoupling corrections to the terms which match on the effective 2HDM but are altered beyond the decoupling limit (for example corrections to the relation between the quark mass and the Yukawa coupling).

\medskip

First we consider the deviations from the 2HDM which are due to imperfect cancellations between the self-energy contributions and the genuine vertex correction. In order to be explicit we consider the coupling of down-quarks to the heavy neutral CP even Higgs (we see from equation \eq{Neutral-Higgs-Squark-vertex} and \eq{Gamma-Squark} that it differs only by factors of $\tan\alpha$, $\tan\beta$ from the other vertices):
\begin{equation}
\renewcommand{\arraystretch}{2.2}
\begin{array}{*{20}c}
   {\left. {\Gamma _{d_f d_t }^{R\,H^0 \,eff} } \right|_{GV}  = \dfrac{2}{{3\pi }}\alpha _s m_{\tilde g} \sum\limits_{s,t = 1}^6 {\sum\limits_{j,k = 1}^3 {\left[ {\dfrac{{e^2 }}{{6c_W^2 }}\left( {v_d \cos \left( \alpha  \right) - v_u \sin \left( \alpha  \right)} \right)\left( {V_{s\,fi}^{LR} \delta _{st}  + \dfrac{{3 - 4s_W^2 }}{{2s_W^2 }}V_{s\,fk}^{LL} V_{t\,ki}^{LR} } \right)} \right.} } } \hfill  \\
   {\;\;\;\;\;\;\;\;\;\;\;\;\;\;\;\;\;\;\;\;\;\;\;\;\;\;\;\;\;\;\;\;\;\;\;\;\; - v_d \left( {Y_k^d } \right)^2 \cos \left( \alpha  \right)\left( {V_{s\,fk}^{d\,LL} V_{t\,ki}^{d\,LR}  + V_{s\,fk}^{d\,LR} V_{t\,ki}^{d\,RR} } \right)} \hfill  \\
   \begin{array}{l}
\;\;\;\;\;\;\;\;\;\;\;\;\;\;\;\;\;\;\;\;\;\;\;\;\;\;\;\;\;\;\;\;\;\;\;\;\; - \tilde \Gamma _d^{H_k^0 } \left[ {\left( {V_{s\,fj}^{d\,LR} A_{kj}^{d * } V_{t\,ki}^{d\,LR}  + V_{s\,fk}^{d\,LL} A_{kj}^d V_{t\,ji}^{d\,RR} } \right)} \right. \\ 
 \left. {\left. {\left. {\;\;\;\;\;\;\;\;\;\;\;\;\;\;\;\;\;\;\;\;\;\;\;\;\;\;\;\;\;\;\;\;\;\;\;\;\; + \tan \left( \alpha  \right)\left( {V_{s\,fj}^{d\,LR} \left( {A_{kj}^{\prime d * }  + \mu ^ *  Y_d^k \delta _{kj} } \right)V_{t\,ki}^{d\,LR}  + V_{s\,fk}^{d\,LL} \left( {A_{kj}^{\prime d}  + \mu Y_d^k \delta _{kj} } \right)V_{t\,ji}^{d\,RR} } \right)} \right)} \right]} \right] \\ 
 \end{array} \hfill  \\
   {\;\;\;\;\;\;\;\;\;\;\;\;\;\;\;\;\;\;\;\;\;\;\;\;\;\; \times \;C_0 \left( {m_{\tilde g}^2 ,m_{\tilde q_t }^2 ,m_{\tilde q_s }^2 } \right) + h.c.} \hfill  \\
\end{array}\label{H-genuine-vertex-corrections}
\end{equation}
The terms in the first two lines vanish in the decoupling limit. In addition there are electroweak contributions which are a priori tiny. Note that the term $V_{s\,fk}^{q\,LL} A_{kj}^q V_{t\,ji}^{q\,RR}$ cancels very precisely with the self-energy correction in the decoupling limit. Therefore, we can only expect sizable decoupling effects from the term proportional to $V_{s\,fj}^{q\,LR} A_{kj}^{q*} V_{t\,ki}^{q\,LR}$. Furthermore, these deviations from an effective 2HDM can only be relevant in the absence of nonholomorphic corrections, because otherwise these corrections also involving the trilinear $A$-terms are dominant. Very large off-diagonal elements $\Delta^{q\,LR}_{ij}$ can, in principle, induce sizable decoupling effects. However, the $A$-terms cannot be arbitrarily large since they are restricted by vacuum stability bounds \cite{Gunion:1987qv,Borzumati:1999sp,Park:2010wf} and 't~Hooft's naturalness criterion \cite{Crivellin:2008mq,Crivellin:2010gw}. Especially the possible size of the off-diagonal elements in the down squark mass matrix  due to $A^d$ cannot be large since $A^d$ enters multiplied with the small vacuum expectation value $v_d$. The combination $v_u A^u$ can be larger, however, their contribution is suppressed by $\cot\beta$. In addition, for the top quark, where one could expect the largest effects due to $A^t$, we have to compare the corrected Yukawa coupling to the huge tree-level one which prohibits sizable effects. In addition, all flavor off-diagonal elements cannot be relevant for decoupling effects since these elements are severely restricted from FCNC processes and must be much smaller than the diagonal ones \footnote{The flavor-changing elements in the up sector can in principle be large, however, their effect on Higgs couplings is again suppressed by $\cot\beta$ and they lead to other dominant non-Higgs-mediated top-FCNCs.}.

\medskip

Therefore, we can only expect relevant decoupling effects from the term $v_d Y^b \mu \tan\beta$ in the down squark mass matrix. In this case the genuine vertex correction is suppressed by $\cot\beta$ and therefore decoupling effects in \eq{H-genuine-vertex-corrections} are irrelevant. However, also the relation between the Yukawa coupling and physical mass in \eq{md-Yd} receives decoupling corrections. These effects can be numerically relevant. In Fig.~\ref{fig:decoupling-corrections} we show the possible effects for SUSY masses of $500\,\rm{GeV}$. We see that only for large negative values of $\mu$ in combination with large $\tan\beta$ sizable effects can occur.

\medskip

\begin{nfigure}{t}
\includegraphics[width=0.8\textwidth]{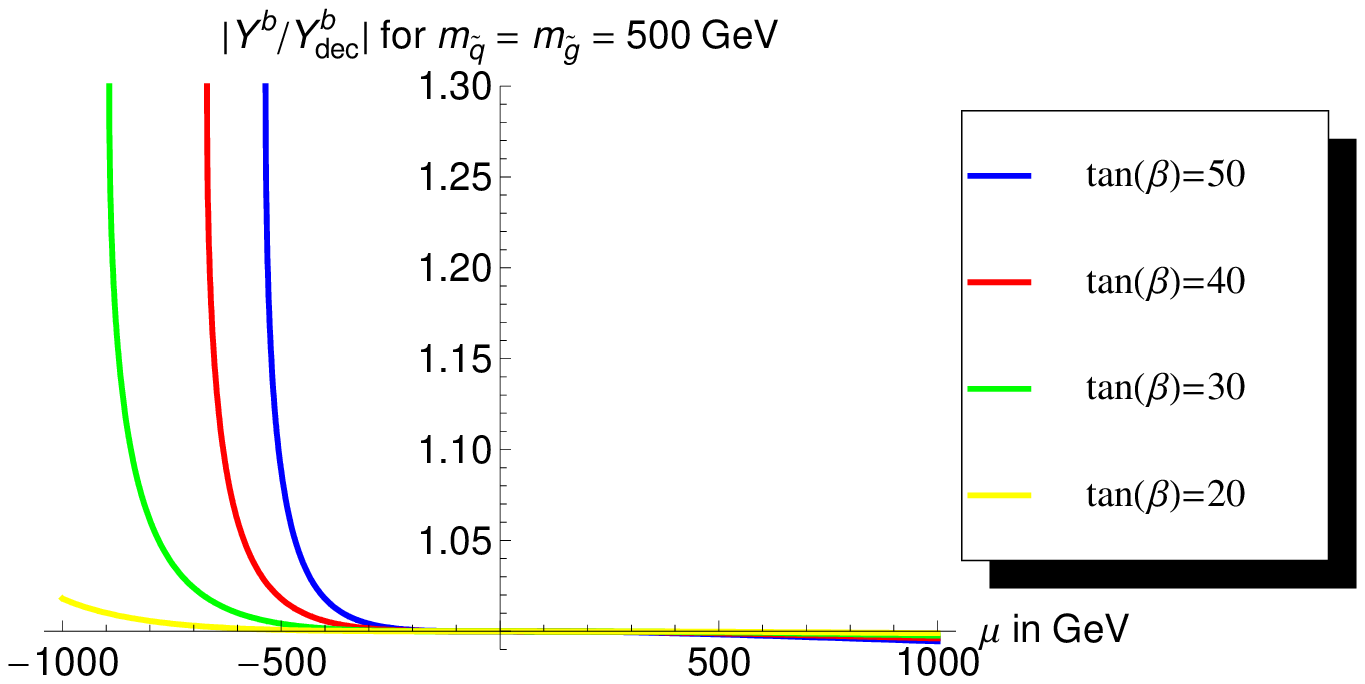}
\caption{Decoupling corrections to the relation between the physical bottom-quark mass and the bottom Yukawa coupling.}\label{fig:decoupling-corrections}
\end{nfigure}

In summary, we conclude that all decoupling effects due to an imperfect cancellation between the genuine vertex correction and the self-energy contributions are negligible. Only the relation between the bottom-quark mass and the Yukawa coupling can receive a sizable decoupling correction for large negative values of the higgsino mass parameter $\mu$. Therefore, the  decoupling limit is an excellent approximation for the full theory, if one uses the non-decoupling relation in order to determine $Y^b$.

\section{Higgs-couplings in the effective-field-theory\label{sec:EFT}}

\begin{nfigure}{t}
\includegraphics[width=0.8\textwidth]{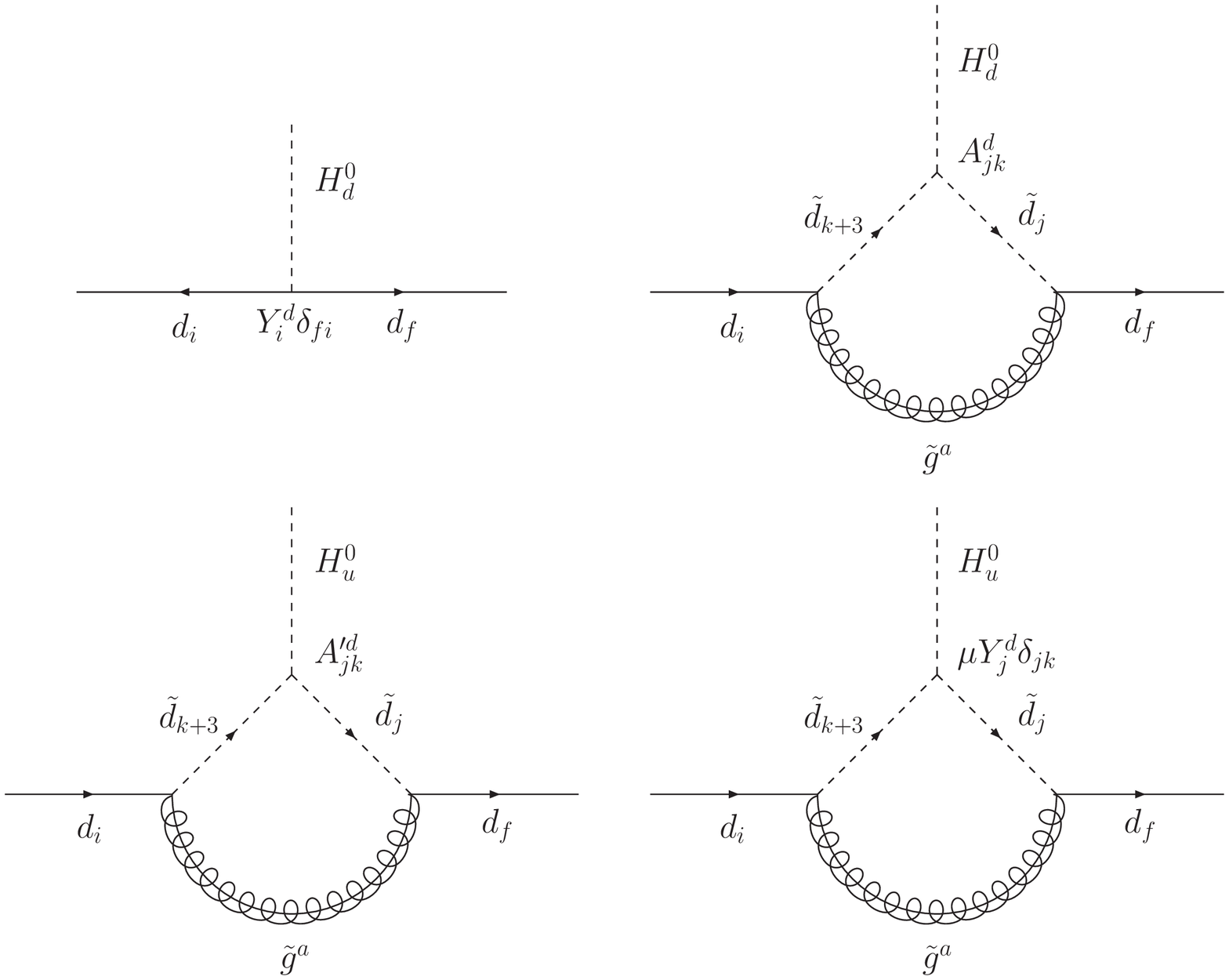}
\caption{Loop-induced Higgs down-quark couplings with gluinos and squarks as virtual particles.}\label{fig:EffectiveHiggsCouplings}
\end{nfigure}

It is instructive to recover the results of the previous section, up to decoupling corrections, using an effective-field-theory approach. This means that if we integrate out the SUSY particles the non-decoupling corrections to Higgs vertices are considered as effective Yukawa couplings. 
Before we specialize to the MSSM let us first consider the most general effective Lagrangian for Yukawa interactions in a 2HDM:
\begin{equation}
L_Y^{eff\;w}  = \bar Q_{f\;L}^a \left( {\left( {E_{fi}^{d\;w}  + Y_{fi}^d } \right)\varepsilon _{ab} H_d^{b\star}  - E_{fi}^{\prime d\;w} H_u^{a} } \right)d_{i\;R}  - \bar Q_{f\;L}^a \left( {\left( {Y_{fi}^u  + E_{fi}^{u\;w} } \right)\varepsilon _{ab} H_u^{b\star}  + E_{fi}^{\prime u\;w} H_d^{a} } \right)u_{i\;R} .
\label{LYeff2HDMw}
\end{equation}
Here a and b denote the components of the SU(2) doublets and $\epsilon_{ab}$ the antisymmetric tensor in two dimensions with $\epsilon_{12}=-1$. We have included nonholomorphic loop-induced corrections $E_{ij}^{\prime d\,w}$ and $E_{ij}^{\prime u\,w}$ as well as the homomorphic ones $E_{ij}^{d\,w}$ and $E_{ij}^{u\,w}$ (the superscript w denotes the fact, that the couplings $E$ in \eq{LYeff2HDMw} are given in a weak basis). The holomorphic corrections were not considered before in the context of effective Higgs couplings in the MSSM. However, they can easily be of order one (of the same size as the quark masses or the corresponding CKM element times the quark mass) \cite{Crivellin:2008mq} and therefore can lead to important effects.
We now decompose the Lagrangian into its neutral and charged interaction part by explicitly writing out the components of the SU(2) doublets and we switch to a basis in which the Yukawa couplings are diagonal in flavor space (this corresponds to the super-CKM basis in the case of the MSSM):
\begin{equation}
\renewcommand{\arraystretch}{2.2}
\begin{array}{l}
 L_Y^{eff}  = \bar u_{f\;L}^{} V_{fk}^{CKM\left( 0 \right)} \left( {\left( {E_{ki}^d  + Y_k^d \delta _{kj} } \right)H_d^{2\star}  - E_{ki}^{\prime d} H_u^{1} }\right)d_{i\;R}  \\ \phantom{ L_Y^{eff}  =} + \bar d_{f\;L}^{} V_{kf}^{CKM\left( 0 \right)*} \left( {\left( {Y_k^u \delta _{ki}  + E_{ki}^u } \right)H_u^{1\star}  - E_{ki}^{\prime u} H_d^{2} } \right)u_{i\;R}  
 \\ \phantom{ L_Y^{eff}  =}  - \bar d_{f\;L}^{} \left( {\left( {E_{fi}^d  + Y_f^d \delta _{fi} } \right)H_d^{1\star}  + E_{fi}^{\prime d} H_u^{2} } \right)d_{i\;R}  
 \\ \phantom{ L_Y^{eff}  =}  - \bar u_{f\;L}^{} \left( {\left( {Y_f^u \delta _{fi}  + E_{fi}^u } \right)H_u^{2\star}  + E_{fi}^{\prime u} H_d^{1} } \right)u_{i\;R}  \\ 
 \end{array}
\end{equation}
Like in the case of the MSSM, $V^{\rm{CKM}\,(0)}$ is the CKM matrix which arises solely due to the misalignment between the (tree-level) Yukawa couplings.
The Higgs fields decompose in the following way into their physical components:
\begin{equation}
\renewcommand{\arraystretch}{2.2}
\begin{array}{l}
 H_d^1  = \dfrac{1}{{\sqrt 2 }}\left( {\cos \left( \alpha  \right)H^0  - \sin \left( \alpha  \right)h^0  + i\sin \left( \beta  \right)A} \right) ,\\ 
 H_u^2  = \dfrac{1}{{\sqrt 2 }}\left( {\sin \left( \alpha  \right)H^0  + \cos \left( \alpha  \right)h^0  + i\cos \left( \beta  \right)A} \right) ,\\ 
 H_u^{1*}  = \cos \left( \beta  \right)H^- ,  \\ 
 H_d^2  = \sin \left( \beta  \right)H^- .  \\ 
 \end{array}
 \label{Higgs-Decomposition}
 \end{equation}
After electroweak symmetry breaking the Higgs fields acquire their vacuum expectation values in their neutral components and the quark mass matrices are afterwards given by:
\begin{equation}
\renewcommand{\arraystretch}{2.2}
\begin{array}{l}
 m^d_{ij}  = \left( {E_{ij}^d  + Y_i^d \delta _{ij} } \right)v_d  + v_u E_{ij}^{\prime d},  \\ 
 m^u_{ij}  = \left( {E_{ij}^u  + Y_i^u \delta _{ij} } \right)v_u  + v_d E_{ij}^{\prime u}.  \\ 
 \end{array}
\end{equation}
The quark mass-matrices are not diagonal in flavor space in this basis (the super-CKM basis in the case of the MSSM) due to the generic corrections $E_{ij}^q$ and $E_{ij}^{\prime q}$. Therefore, we have to diagonalize them by a bi-unitary transformation
\begin{equation}
U^{q\,L\dag } m^q U^{q\,R}  = m_{q_i } 
\label{diagonalization}
\end{equation}
in order to receive the physical quark masses $m_{q_i}$. We can assume that the off-diagonal entries are smaller than the diagonal ones since we know from experiment that flavor violation is a small quantity \footnote{Since the rotation matrices $U^{q\,L,R}$ are not physical this assumption relies implicitly also on a naturalness criterion which prohibits large cancellations.}. In addition, if the corrections $E_{ij}^q$ and $E_{ij}^{\prime q}$ are loop-induced, at least the top Yukawa coupling is bigger that the off-diagonal entries. Therefore, it is possible to perform a perturbative diagonalization. We get for the following rotation matrices which diagonalize the down-quark mass matrix:
\begin{equation}
\renewcommand{\arraystretch}{2.2}
\begin{array}{l}
 U_{}^{d\,L}  = \left( {\begin{array}{*{20}c}
   1 & {\dfrac{{E_{12}^d v_d  + E_{12}^{\prime d} v_u }}{{m_{d_2 } }}} & {\dfrac{{E_{13}^d v_d  + E_{13}^{\prime d} v_u }}{{m_{d_3 } }}}  \\
   {\dfrac{{E_{12}^{d*} v_d  + E_{12}^{\prime d*} v_u }}{{m_{d_2 } }}} & 1 & {\dfrac{{E_{23}^d v_d  + E_{23}^{\prime d} v_u }}{{m_{d_3 } }}}  \\
   {\dfrac{{E_{13}^{d*} v_d  + E_{13}^{\prime d*} v_u }}{{m_{d_3 } }}} & {\dfrac{{E_{23}^{d*} v_d  + E_{23}^{\prime d*} v_u }}{{m_{d_3 } }}} & 1  \\
\end{array}} \right), \\ 
 U_{}^{d\,R}  = \left( {\begin{array}{*{20}c}
   1 & {\dfrac{{E_{21}^d v_d  + E_{21}^{\prime d} v_u }}{{m_{d_2 } }}} & {\dfrac{{E_{31}^d v_d  + E_{31}^{\prime d} v_u }}{{m_{d_3 } }}}  \\
   {\dfrac{{E_{21}^{d*} v_d  + E_{21}^{\prime d*} v_u }}{{m_{d_2 } }}} & 1 & {\dfrac{{E_{32}^d v_d  + E_{32}^{\prime d} v_u }}{{m_{d_3 } }}}  \\
   {\dfrac{{E_{31}^{d*} v_d  + E_{31}^{\prime d*} v_u }}{{m_{d_3 } }}} & {\dfrac{{E_{32}^{d*} v_d  + E_{32}^{\prime d*} v_u }}{{m_{d_3 } }}} & 1  \\
\end{array}} \right). \\ 
 \end{array}
 \label{Ueff}
\end{equation}
For up-quarks the rotation matrices are simply obtained by interchanging $u$ and $d$. At leading order, the masses $m_{q_i}$ are just given by the diagonal entries of the mass matrix:
\begin{equation}
\renewcommand{\arraystretch}{2.2}
\begin{array}{l}
 m_{d_i }  = \left( {E_{ii}^d  + Y_i^d } \right)v_d  + v_u E_{ii}^{\prime d},  \\ 
 m_{u_i }  = \left( {E_{ii}^u  + Y_i^u } \right)v_u  + v_d E_{ii}^{\prime u}.  \\ 
 \end{array}
 \label{mass-eigenvalues}
\end{equation}
Now, in order to calculate the flavor structure of the Higgs Yukawa couplings we switch to the physical basis in which the quark masses are diagonal:
\begin{equation}
\renewcommand{\arraystretch}{2.2}
\begin{array}{l}
 L_Y^{eff}  = \bar u_{f\;L}^{} U_{kf}^{u\;L*} \left( {\left( {E_{kj}^d  + Y_j^d \delta _{kj} } \right)H_d^{2\star}  - E_{kj}^{\prime d} H_u^{1} } \right)U_{ji}^{d\;R} d_{i\;R}  
 \\ \phantom{ L_Y^{eff}  =}+ \bar d_{f\;L}^{} U_{kf}^{d\;L*} \left( {\left( {Y_j^u \delta _{kj}  + E_{kj}^u } \right)H_u^{1\star}  - E_{kj}^{\prime u} H_d^{2} } \right)U_{ji}^{u\;R} u_{i\;R}  
 \\ \phantom{ L_Y^{eff}  =} - \bar d_{f\;L}^{} U_{kf}^{d\;L*} \left( {\left( {E_{kj}^d  + Y_j^d \delta _{kj} } \right)H_d^{1\star}  + E_{kj}^{\prime d} H_u^{2} } \right)U_{ji}^{d\;R} d_{i\;R} 
  \\ \phantom{ L_Y^{eff}  =} - \bar u_{f\;L}^{} U_{kf}^{u\;L*} \left( {\left( {Y_j^u \delta _{kj}  + E_{kj}^u } \right)H_u^{2\star}  + E_{kj}^{\prime u} H_d^{1} } \right)U_{ji}^{u\;R} u_{i\;R} \\ 
 \end{array}
\end{equation}
We now eliminate the explicit dependence on the bare Yukawa couplings $Y^q$ by using ~\eq{mass-eigenvalues} and \eq{diagonalization}:
\begin{equation}
\renewcommand{\arraystretch}{2.2}
\begin{array}{l}
 L_Y^{eff}  =  - \bar d_{f\;L} \left[ {-\tilde H^0 \tan \left( \beta  \right)\tilde E_{fi}^{\prime d}  + H_d^1 \dfrac{{m_{d_i } }}{{v_d }}\delta _{fi} } \right]d_{i\;R}  \\ 
  \phantom{L_Y^{eff}  =  } - \bar u_{f\;L} \left[ {\tilde H^{0*} \tilde E_{fi}^{\prime u}  + H_u^2 \dfrac{{m_{u_i } }}{{v_u }}\delta _{fi} } \right]u_{i\;R}  \\ 
  \phantom{L_Y^{eff}  =  } + \bar u_{f\;L} V_{fj}^{CKM} \left[ {-\left( {\cot \left( \beta  \right) + \tan \left( \beta  \right)} \right)\tilde E_{ji}^{\prime d}  + \dfrac{{m_{d_i } }}{{v_d }}\delta _{ji} } \right]\sin \left( \beta  \right)H^ +  d_{i\;R}  \\ 
  \phantom{L_Y^{eff}  =  } + \bar d_{f\;L} V_{jf}^{CKM*} \left[ {-\left( {\tan \left( \beta  \right) + \cot \left( \beta  \right)} \right)\tilde E_{ji}^{\prime u}  + \dfrac{{m_{u_i } }}{{v_u }}\delta _{ji} } \right]\cos \left( \beta  \right)H^ -  u_{i\;R}  \\ 
 \end{array}
 \label{L-Y-FCNC}
\end{equation}
with
\begin{equation}
 \tilde E_{fi}^{\prime q}  = U_{kf}^{q\,L*} E_{kj}^{\prime q} U_{ji}^{q\,R} , 
 \end{equation}
 \begin{equation}
 \tilde H^0  = {\cot \left( \beta  \right)H_u^{2*}  - H_d^1 }  = 
   {\dfrac{{\sin \left( {\alpha  - \beta } \right)}}{{\sqrt 2 \sin \left( \beta  \right)}}H^0 } + {\dfrac{{\cos \left( {\alpha  - \beta } \right)}}{{\sqrt 2 \sin \left( \beta  \right)}}h^0 } - {\dfrac{{ i}}{{\sqrt 2 }}\sin \left( \beta  \right)A^0 } . \label{Htilde} 
 \end{equation}
All FCNC interactions are contained in the first terms in the first two lines of \eq{L-Y-FCNC} while the second terms in the first two lines are flavor-diagonal and reproduce the tree-level Higgs couplings in the absence of loop corrections. 
We can evaluate the terms $E_{fi}^{\prime q}$ with the help of \eq{Ueff}. It is only assumed that the off-diagonal terms are smaller than the diagonal ones, but we treat the flavor-diagonal terms as order one corrections (which indeed can be the case for down-type quarks). However, the flavor-diagonal corrections are assumed to posses the same hierarchy as the quark masses. If we further neglect small mass rations we get:
\begin{equation}
\renewcommand{\arraystretch}{2.2}
\tilde E_{fi}^{\prime d}  = E_{fi}^{\prime d}  - \left( {\begin{array}{*{20}c}
   0 & {\dfrac{{E_{22}^{\prime d} \left( {E_{12}^d v_d  + E_{12}^{\prime d} v_u } \right)}}{{m_{d_2 } }}} & {\dfrac{{E_{33}^{\prime d} \left( {E_{13}^d v_d  + E_{13}^{\prime d} v_u } \right)}}{{m_{d_3 } }}}  \\
   {\dfrac{{E_{22}^{\prime d} \left( {E_{21}^d v_d  + E_{21}^{\prime d} v_u } \right)}}{{m_{d_2 } }}} & 0 & {\dfrac{{E_{33}^{\prime d} \left( {E_{23}^d v_d  + E_{23}^{\prime d} v_u } \right)}}{{m_{d_3 } }}}  \\
   {\dfrac{{E_{33}^{\prime d} \left( {E_{31}^d v_d  + E_{31}^{\prime d} v_u } \right)}}{{m_{d_3 } }}} & {\dfrac{{E_{33}^{\prime d} \left( {E_{31}^d v_d  + E_{31}^{\prime d} v_u } \right)}}{{m_{d_3 } }}} & 0  \\
\end{array}} \right)_{fi}
\label{eq:Etilde}
\end{equation}
The expression for $E_{fi}^{\prime u}$ is obtained by simply exchanging u and d.

We now want to evaluate \eq{L-Y-FCNC} for the special case of the MSSM where we have as discussed in the previous section:
\begin{equation}
m^q_{ij}  = v_q Y_i^q \delta _{ij}  + \Sigma _{ij}^{q\,LR} 
\end{equation}
The self-energy contribution  $\Sigma _{ij}^{q\,LR}$ decomposes according to \eq{eq:self-energy-decomposition}. However, since we work in the decoupling limit, the parts of the self-energy $\Sigma _{fi\,A}^{q\,LR}$, $\Sigma _{fi\,A^{\prime}}^{q\,LR}$ and $\Sigma _{fi\,Y^q_j}^{q\,LR}$ are now linear in $A$, $A^{\prime}$ and $Y^q_j$, respectively. This means that the combinations of rotation matrices $V_{s\,ij}^{q\,LL,RR}$ depend only on the bilinear terms of the squark mass matrices. Therefore, the corrections $E_{ij}^q ,\,E_{ij}^{\prime q}$ shown in Fig.~\ref{fig:EffectiveHiggsCouplings} are 
\begin{equation}
\renewcommand{\arraystretch}{2.2}
\begin{array}{*{20}c}
   {E_{ij}^d  = \dfrac{{\Sigma _{ij\,A}^{d\,LR} }}{{v_d }},}\qquad & {E_{ij}^{\prime d}  = \dfrac{{\Sigma _{ij\,A^\prime}^{d\,LR} }}{{v_u }} + \varepsilon _i^d Y_i^d \delta _{ij} },  \\
   {E_{ij}^u  = \dfrac{{\Sigma _{ij\,A}^{u\,LR} }}{{v_u }},} \qquad & {E_{ij}^{\prime u}  = \dfrac{{\Sigma _{ij\,A^\prime}^{u\,LR} }}{{v_d }}},  \\
\end{array}
\end{equation}
with $\varepsilon _i^d$ defined in \eq{eq:epsilon_d}. Therefore, \eq{eq:Etilde} is simplify given by
\begin{eqnarray}
\tilde E_{fi}^{\prime u}  = \dfrac{1}{{v_u }}\tan \left( \beta  \right)\tilde \Sigma _{fi\,A^\prime }^{u\,LR},\\
\tilde E_{fi}^{\prime d}  = \frac{1}{{v_d }}\cot \left( \beta  \right)\tilde \Sigma _{fi\,A^\prime \mu }^{d\,LR} , 
\end{eqnarray}
with $\tilde \Sigma _{fi}^{q\,LR}$ defined in \eq{Sigma-tilde}. If we plug this into \eq{L-Y-FCNC} we receive:
\begin{equation}
\renewcommand{\arraystretch}{2.2}
\begin{array}{l}
 L_Y^{eff}  =  - \bar d_{f\;L} \dfrac{1}{{v_d }}\left[ {-\tilde H^0 \tilde \Sigma _{fi\;A'\mu }^{d\;LR}  + H_d^1 m_{d_i } \delta _{fi} } \right]d_{i\;R}  \\ 
  \phantom{L_Y^{eff}  =}  - \bar u_{f\;L} \dfrac{1}{{v_u }}\left[ {\tilde H^{0*} \tan \left( \beta  \right)\tilde \Sigma _{fi\;A'}^{u\;LR}  + H_u^2 m_{u_i } \delta _{fi} } \right]u_{i\;R}  \\ 
 \phantom{L_Y^{eff}  =}  + \bar u_{f\;L} V_{fj}^{CKM} \dfrac{1}{v}\left[ {-\left( {\cot \left( \beta  \right) + \tan \left( \beta  \right)} \right)\tilde \Sigma _{ji\;A'\mu }^{d\;LR}  + \tan \left( \beta  \right)m_{d_i } \delta _{ji} } \right]H^ +  d_{i\;R}  \\ 
 \phantom{L_Y^{eff}  =}  + \bar d_{f\;L} V_{jf}^{CKM*} \dfrac{1}{v}\left[ {-\left( {\tan \left( \beta  \right) + \cot \left( \beta  \right)} \right)\tilde \Sigma _{ji\;A'}^{u\;LR}  + \cot \left( \beta  \right)m_{u_i } \delta _{ji} } \right]H^ -  u_{i\;R}  \\ 
 \end{array}
\label{higgs-couplings-EFT}
\end{equation}
If we multiply this expression by $i$ and plug in the decomposition of the Higgs field in \eq{Higgs-Decomposition} and \eq{Htilde} we recover the result obtained in the decoupling limit of the full theory given in \eq{Higgs-vertices-decoupling}.

\medskip

Comparing our results in \eq{higgs-couplings-EFT} or \eq{Higgs-vertices-decoupling} with the literature \cite{Hamzaoui:1998nu,Babu:1999hn,Chankowski:2000ng,Isidori:2002qe,Dedes:2008iw} we find new contributions proportional $\Sigma^{q\,LR}_{ij}\Sigma^{q\,LR}_{\rm{max}(i,j)\,A^\prime \mu}/m_{q_{\rm{max}(i,j)}}$. These terms are numerically important for down-quarks if $\tan\beta$ is large, since both $\Sigma^{d\,LR}_{ij}$ and $\Sigma^{d\,LR}_{ii}$ can be of order one compared to $V^{\rm{CKM}}_{ij}/m_{d_{\rm{max}(i,j)}}$ or $m_{d_i}$, respectively. We obtain these new terms for two reasons: First we did not neglect the holomorphic corrections $E_{ij}^q$. Second, in applying the rotations in flavor-space $U^{q\,L.R}$ which diagonalize the masses, we kept terms proportional to  $E_{ij}^q E_{\rm{max(i,j)}}^{\prime q}$ or $E_{ij}^{\prime q} E_{\rm{max(i,j)}}^{\prime q}$. This is consistent with our diagrammatic approach in Sec.~\ref{sec:Renormalization} where we found that two-loop corrections were necessary in order to get the correct result in the decoupling limit.

\section{conclusions\label{sec:conclusions}}

In this article we have calculated the effective quark-Higgs couplings, including the corrections from squark-gluino diagrams \footnote{An analysis including also the electroweak contributions is forthcoming \cite{Crivellin:2011}.}, in the most general MSSM. We have performed our calculations using an purely diagrammatic approach in a minimal renormalization scheme which simplifies the resummation of $\tan\beta$. In order to include correctly all chirally enhanced effects one is forced to consider diagrams which are formally of order $\alpha_s^2$. We confirm this statement in the effective-field-theory approach whose result is only obtained in the full theory if the flavor-changing wave-function rotation induced by the self-energies is also applied to the genuine vertex correction.

\medskip

In subsection \ref{sec:decoupling-corrections} we have addressed the issue of decoupling corrections to the effective Higgs vertices. It turns out that the decoupling limit excellently reproduces the full result apart from possible corrections to the relation between the bottom-quark mass and the bottom Yukawa coupling $Y^b$ which can be relevant for large negative values of $\mu$. However, in order to include these corrections it is sufficient to use the result obtained in the decoupling limit with the value of $Y^b$ calculated to all orders in $v/M_{\rm{SUSY}}$.

\medskip

We have found that the $A'$-terms induce flavor-changing neutral Higgs coupling similar the known effects stemming from nonholomorphic corrections involving $\tan\beta$, but with a generic flavor structure. Therefore, in the decoupling limit the MSSM could, in principle, lead to a general 2HDM of type III with all interesting flavor effects present in this model.

\medskip

In addition, we have found new $\tan\beta$ enhanced corrections which were not discussed before in the literature. From \eq{Higgs-vertices-decoupling} and \eq{Sigma-tilde} we see that these contributions are proportional to the product $\Sigma^{q\,LR}_{ij}\Sigma^{q\,LR}_{\rm{max}(i,j)\,A^\prime \mu}/m_{d_{\rm{max}(i,j)}}$. Because of a chiral enhancement, both $\Sigma^{d\,LR}_{ij}$ and $\Sigma^{d\,LR}_{ii}$ can be of order one, compared to $V^{\rm{CKM}}_{ij}/m_{d_{\rm{max}(i,j)}}$ or $m_{d_i}$, respectively. Therefore, these new contributions can be numerically highly relevant.

\medskip

In Sec.~\ref{sec:EFT} we have calculated the effective Higgs couplings using an effective-field-theory approach. We included both holomorphic and nonholomorphic corrections to the Higgs coupling. With these corrections, the Higgs couplings are no longer diagonal in the same basis as quark mass matrices which leads to flavor-changing Higgs vertices. We then specified to the MSSM and recovered the result obtained in the decoupling limit of the full theory. This also verifies our statement that chirally enhanced self-energies are physical and cannot be renormalized away once also Higgs mediated processes are considered.

\vspace{7mm}
{\it Acknowledgments.}--- I am grateful to Ulrich Nierste, Christoph Greub, Jernej Kamenik and Janusz Rosiek for useful discussions and proofreading the article. I also like to thank Lars Hofer for useful discussion concerning the different renormalization schemes.
A.C.~is supported by the Swiss
National Foundation.  The Albert Einstein Center for Fundamental Physics
is supported by the ``Innovations- und Kooperationsprojekt C-13 of the
Schweizerische Universit\"atskonferenz SUK/CRUS''.

\bibliography{Effective-Higgs-Vertices}

\begin{thebibliography}{32}
\expandafter\ifx\csname natexlab\endcsname\relax\def\natexlab#1{#1}\fi
\expandafter\ifx\csname bibnamefont\endcsname\relax
  \def\bibnamefont#1{#1}\fi
\expandafter\ifx\csname bibfnamefont\endcsname\relax
  \def\bibfnamefont#1{#1}\fi
\expandafter\ifx\csname citenamefont\endcsname\relax
  \def\citenamefont#1{#1}\fi
\expandafter\ifx\csname url\endcsname\relax
  \def\url#1{\texttt{#1}}\fi
\expandafter\ifx\csname urlprefix\endcsname\relax\def\urlprefix{URL }\fi
\providecommand{\bibinfo}[2]{#2}
\providecommand{\eprint}[2][]{\url{#2}}

\bibitem[{\citenamefont{Hamzaoui et~al.}(1999)\citenamefont{Hamzaoui, Pospelov,
  and Toharia}}]{Hamzaoui:1998nu}
\bibinfo{author}{\bibfnamefont{C.}~\bibnamefont{Hamzaoui}},
  \bibinfo{author}{\bibfnamefont{M.}~\bibnamefont{Pospelov}}, \bibnamefont{and}
  \bibinfo{author}{\bibfnamefont{M.}~\bibnamefont{Toharia}},
  \bibinfo{journal}{Phys. Rev.} \textbf{\bibinfo{volume}{D59}},
  \bibinfo{pages}{095005} (\bibinfo{year}{1999}), \eprint{hep-ph/9807350}.

\bibitem[{\citenamefont{Babu and Kolda}(2000)}]{Babu:1999hn}
\bibinfo{author}{\bibfnamefont{K.~S.} \bibnamefont{Babu}} \bibnamefont{and}
  \bibinfo{author}{\bibfnamefont{C.~F.} \bibnamefont{Kolda}},
  \bibinfo{journal}{Phys. Rev. Lett.} \textbf{\bibinfo{volume}{84}},
  \bibinfo{pages}{228} (\bibinfo{year}{2000}), \eprint{hep-ph/9909476}.

\bibitem[{\citenamefont{Isidori and Retico}(2001)}]{Isidori:2001fv}
\bibinfo{author}{\bibfnamefont{G.}~\bibnamefont{Isidori}} \bibnamefont{and}
  \bibinfo{author}{\bibfnamefont{A.}~\bibnamefont{Retico}},
  \bibinfo{journal}{JHEP} \textbf{\bibinfo{volume}{11}}, \bibinfo{pages}{001}
  (\bibinfo{year}{2001}), \eprint{hep-ph/0110121}.

\bibitem[{\citenamefont{Chankowski and Slawianowska}(2001)}]{Chankowski:2000ng}
\bibinfo{author}{\bibfnamefont{P.~H.} \bibnamefont{Chankowski}}
  \bibnamefont{and}
  \bibinfo{author}{\bibfnamefont{L.}~\bibnamefont{Slawianowska}},
  \bibinfo{journal}{Phys. Rev.} \textbf{\bibinfo{volume}{D63}},
  \bibinfo{pages}{054012} (\bibinfo{year}{2001}), \eprint{hep-ph/0008046}.

\bibitem[{\citenamefont{Buras et~al.}(2002)\citenamefont{Buras, Chankowski,
  Rosiek, and Slawianowska}}]{Buras:2002wq}
\bibinfo{author}{\bibfnamefont{A.~J.} \bibnamefont{Buras}},
  \bibinfo{author}{\bibfnamefont{P.~H.} \bibnamefont{Chankowski}},
  \bibinfo{author}{\bibfnamefont{J.}~\bibnamefont{Rosiek}}, \bibnamefont{and}
  \bibinfo{author}{\bibfnamefont{L.}~\bibnamefont{Slawianowska}},
  \bibinfo{journal}{Phys. Lett.} \textbf{\bibinfo{volume}{B546}},
  \bibinfo{pages}{96} (\bibinfo{year}{2002}), \eprint{hep-ph/0207241}.

\bibitem[{\citenamefont{Buras et~al.}(2003)\citenamefont{Buras, Chankowski,
  Rosiek, and Slawianowska}}]{Buras:2002vd}
\bibinfo{author}{\bibfnamefont{A.~J.} \bibnamefont{Buras}},
  \bibinfo{author}{\bibfnamefont{P.~H.} \bibnamefont{Chankowski}},
  \bibinfo{author}{\bibfnamefont{J.}~\bibnamefont{Rosiek}}, \bibnamefont{and}
  \bibinfo{author}{\bibfnamefont{L.}~\bibnamefont{Slawianowska}},
  \bibinfo{journal}{Nucl. Phys.} \textbf{\bibinfo{volume}{B659}},
  \bibinfo{pages}{3} (\bibinfo{year}{2003}), \eprint{hep-ph/0210145}.

\bibitem[{\citenamefont{Dedes et~al.}(2001)\citenamefont{Dedes, Dreiner, and
  Nierste}}]{Dedes:2001fv}
\bibinfo{author}{\bibfnamefont{A.}~\bibnamefont{Dedes}},
  \bibinfo{author}{\bibfnamefont{H.~K.} \bibnamefont{Dreiner}},
  \bibnamefont{and} \bibinfo{author}{\bibfnamefont{U.}~\bibnamefont{Nierste}},
  \bibinfo{journal}{Phys. Rev. Lett.} \textbf{\bibinfo{volume}{87}},
  \bibinfo{pages}{251804} (\bibinfo{year}{2001}), \eprint{hep-ph/0108037}.

\bibitem[{\citenamefont{Bobeth et~al.}(2002)\citenamefont{Bobeth, Ewerth,
  Kruger, and Urban}}]{Bobeth:2002ch}
\bibinfo{author}{\bibfnamefont{C.}~\bibnamefont{Bobeth}},
  \bibinfo{author}{\bibfnamefont{T.}~\bibnamefont{Ewerth}},
  \bibinfo{author}{\bibfnamefont{F.}~\bibnamefont{Kruger}}, \bibnamefont{and}
  \bibinfo{author}{\bibfnamefont{J.}~\bibnamefont{Urban}},
  \bibinfo{journal}{Phys. Rev.} \textbf{\bibinfo{volume}{D66}},
  \bibinfo{pages}{074021} (\bibinfo{year}{2002}), \eprint{hep-ph/0204225}.

\bibitem[{\citenamefont{Baek et~al.}(2002)\citenamefont{Baek, Ko, and
  Song}}]{Baek:2002rt}
\bibinfo{author}{\bibfnamefont{S.}~\bibnamefont{Baek}},
  \bibinfo{author}{\bibfnamefont{P.}~\bibnamefont{Ko}}, \bibnamefont{and}
  \bibinfo{author}{\bibfnamefont{W.~Y.} \bibnamefont{Song}},
  \bibinfo{journal}{Phys. Rev. Lett.} \textbf{\bibinfo{volume}{89}},
  \bibinfo{pages}{271801} (\bibinfo{year}{2002}), \eprint{hep-ph/0205259}.

\bibitem[{\citenamefont{Mizukoshi et~al.}(2002)\citenamefont{Mizukoshi, Tata,
  and Wang}}]{Mizukoshi:2002gs}
\bibinfo{author}{\bibfnamefont{J.~K.} \bibnamefont{Mizukoshi}},
  \bibinfo{author}{\bibfnamefont{X.}~\bibnamefont{Tata}}, \bibnamefont{and}
  \bibinfo{author}{\bibfnamefont{Y.}~\bibnamefont{Wang}},
  \bibinfo{journal}{Phys. Rev.} \textbf{\bibinfo{volume}{D66}},
  \bibinfo{pages}{115003} (\bibinfo{year}{2002}), \eprint{hep-ph/0208078}.

\bibitem[{\citenamefont{Dedes et~al.}(2009)\citenamefont{Dedes, Rosiek, and
  Tanedo}}]{Dedes:2008iw}
\bibinfo{author}{\bibfnamefont{A.}~\bibnamefont{Dedes}},
  \bibinfo{author}{\bibfnamefont{J.}~\bibnamefont{Rosiek}}, \bibnamefont{and}
  \bibinfo{author}{\bibfnamefont{P.}~\bibnamefont{Tanedo}},
  \bibinfo{journal}{Phys. Rev.} \textbf{\bibinfo{volume}{D79}},
  \bibinfo{pages}{055006} (\bibinfo{year}{2009}), \eprint{0812.4320}.

\bibitem[{\citenamefont{Dedes and Pilaftsis}(2003)}]{Dedes:2002er}
\bibinfo{author}{\bibfnamefont{A.}~\bibnamefont{Dedes}} \bibnamefont{and}
  \bibinfo{author}{\bibfnamefont{A.}~\bibnamefont{Pilaftsis}},
  \bibinfo{journal}{Phys. Rev.} \textbf{\bibinfo{volume}{D67}},
  \bibinfo{pages}{015012} (\bibinfo{year}{2003}), \eprint{hep-ph/0209306}.

\bibitem[{\citenamefont{Carena et~al.}(2006)\citenamefont{Carena, Menon,
  Noriega-Papaqui, Szynkman, and Wagner}}]{Carena:2006ai}
\bibinfo{author}{\bibfnamefont{M.~S.} \bibnamefont{Carena}},
  \bibinfo{author}{\bibfnamefont{A.}~\bibnamefont{Menon}},
  \bibinfo{author}{\bibfnamefont{R.}~\bibnamefont{Noriega-Papaqui}},
  \bibinfo{author}{\bibfnamefont{A.}~\bibnamefont{Szynkman}}, \bibnamefont{and}
  \bibinfo{author}{\bibfnamefont{C.~E.~M.} \bibnamefont{Wagner}},
  \bibinfo{journal}{Phys. Rev.} \textbf{\bibinfo{volume}{D74}},
  \bibinfo{pages}{015009} (\bibinfo{year}{2006}), \eprint{hep-ph/0603106}.

\bibitem[{\citenamefont{Hall et~al.}(1994)\citenamefont{Hall, Rattazzi, and
  Sarid}}]{Hall:1993gn}
\bibinfo{author}{\bibfnamefont{L.~J.} \bibnamefont{Hall}},
  \bibinfo{author}{\bibfnamefont{R.}~\bibnamefont{Rattazzi}}, \bibnamefont{and}
  \bibinfo{author}{\bibfnamefont{U.}~\bibnamefont{Sarid}},
  \bibinfo{journal}{Phys. Rev.} \textbf{\bibinfo{volume}{D50}},
  \bibinfo{pages}{7048} (\bibinfo{year}{1994}), \eprint{hep-ph/9306309}.

\bibitem[{\citenamefont{Blazek et~al.}(1995)\citenamefont{Blazek, Raby, and
  Pokorski}}]{Blazek:1995nv}
\bibinfo{author}{\bibfnamefont{T.}~\bibnamefont{Blazek}},
  \bibinfo{author}{\bibfnamefont{S.}~\bibnamefont{Raby}}, \bibnamefont{and}
  \bibinfo{author}{\bibfnamefont{S.}~\bibnamefont{Pokorski}},
  \bibinfo{journal}{Phys. Rev.} \textbf{\bibinfo{volume}{D52}},
  \bibinfo{pages}{4151} (\bibinfo{year}{1995}), \eprint{hep-ph/9504364}.

\bibitem[{\citenamefont{Hofer et~al.}(2009)\citenamefont{Hofer, Nierste, and
  Scherer}}]{Hofer:2009xb}
\bibinfo{author}{\bibfnamefont{L.}~\bibnamefont{Hofer}},
  \bibinfo{author}{\bibfnamefont{U.}~\bibnamefont{Nierste}}, \bibnamefont{and}
  \bibinfo{author}{\bibfnamefont{D.}~\bibnamefont{Scherer}},
  \bibinfo{journal}{JHEP} \textbf{\bibinfo{volume}{10}}, \bibinfo{pages}{081}
  (\bibinfo{year}{2009}), \eprint{0907.5408}.

\bibitem[{\citenamefont{Isidori and Retico}(2002)}]{Isidori:2002qe}
\bibinfo{author}{\bibfnamefont{G.}~\bibnamefont{Isidori}} \bibnamefont{and}
  \bibinfo{author}{\bibfnamefont{A.}~\bibnamefont{Retico}},
  \bibinfo{journal}{JHEP} \textbf{\bibinfo{volume}{09}}, \bibinfo{pages}{063}
  (\bibinfo{year}{2002}), \eprint{hep-ph/0208159}.

\bibitem[{\citenamefont{Crivellin and Nierste}(2009)}]{Crivellin:2008mq}
\bibinfo{author}{\bibfnamefont{A.}~\bibnamefont{Crivellin}} \bibnamefont{and}
  \bibinfo{author}{\bibfnamefont{U.}~\bibnamefont{Nierste}},
  \bibinfo{journal}{Phys. Rev.} \textbf{\bibinfo{volume}{D79}},
  \bibinfo{pages}{035018} (\bibinfo{year}{2009}), \eprint{0810.1613}.

\bibitem[{\citenamefont{Crivellin and Nierste}(2010)}]{Crivellin:2009ar}
\bibinfo{author}{\bibfnamefont{A.}~\bibnamefont{Crivellin}} \bibnamefont{and}
  \bibinfo{author}{\bibfnamefont{U.}~\bibnamefont{Nierste}},
  \bibinfo{journal}{Phys. Rev.} \textbf{\bibinfo{volume}{D81}},
  \bibinfo{pages}{095007} (\bibinfo{year}{2010}), \eprint{0908.4404}.

\bibitem[{\citenamefont{Hall and Randall}(1990)}]{Hall:1990ac}
\bibinfo{author}{\bibfnamefont{L.~J.} \bibnamefont{Hall}} \bibnamefont{and}
  \bibinfo{author}{\bibfnamefont{L.}~\bibnamefont{Randall}},
  \bibinfo{journal}{Phys. Rev. Lett.} \textbf{\bibinfo{volume}{65}},
  \bibinfo{pages}{2939} (\bibinfo{year}{1990}).

\bibitem[{\citenamefont{Haber and Mason}(2008)}]{Haber:2007dj}
\bibinfo{author}{\bibfnamefont{H.~E.} \bibnamefont{Haber}} \bibnamefont{and}
  \bibinfo{author}{\bibfnamefont{J.~D.} \bibnamefont{Mason}},
  \bibinfo{journal}{Phys. Rev.} \textbf{\bibinfo{volume}{D77}},
  \bibinfo{pages}{115011} (\bibinfo{year}{2008}), \eprint{0711.2890}.

\bibitem[{\citenamefont{Rosiek}(1990)}]{Rosiek:1989rs}
\bibinfo{author}{\bibfnamefont{J.}~\bibnamefont{Rosiek}},
  \bibinfo{journal}{Phys. Rev.} \textbf{\bibinfo{volume}{D41}},
  \bibinfo{pages}{3464} (\bibinfo{year}{1990}).

\bibitem[{\citenamefont{Rosiek}(1995)}]{Rosiek:1995kg}
\bibinfo{author}{\bibfnamefont{J.}~\bibnamefont{Rosiek}}
  (\bibinfo{year}{1995}), \eprint{hep-ph/9511250}.

\bibitem[{\citenamefont{Logan and Nierste}(2000)}]{Logan:2000iv}
\bibinfo{author}{\bibfnamefont{H.~E.} \bibnamefont{Logan}} \bibnamefont{and}
  \bibinfo{author}{\bibfnamefont{U.}~\bibnamefont{Nierste}},
  \bibinfo{journal}{Nucl. Phys.} \textbf{\bibinfo{volume}{B586}},
  \bibinfo{pages}{39} (\bibinfo{year}{2000}), \eprint{hep-ph/0004139}.

\bibitem[{\citenamefont{Carena et~al.}(2000)\citenamefont{Carena, Garcia,
  Nierste, and Wagner}}]{Carena:1999py}
\bibinfo{author}{\bibfnamefont{M.~S.} \bibnamefont{Carena}},
  \bibinfo{author}{\bibfnamefont{D.}~\bibnamefont{Garcia}},
  \bibinfo{author}{\bibfnamefont{U.}~\bibnamefont{Nierste}}, \bibnamefont{and}
  \bibinfo{author}{\bibfnamefont{C.~E.~M.} \bibnamefont{Wagner}},
  \bibinfo{journal}{Nucl. Phys.} \textbf{\bibinfo{volume}{B577}},
  \bibinfo{pages}{88} (\bibinfo{year}{2000}), \eprint{hep-ph/9912516}.

\bibitem[{\citenamefont{Crivellin and Girrbach}(2010)}]{Crivellin:2010gw}
\bibinfo{author}{\bibfnamefont{A.}~\bibnamefont{Crivellin}} \bibnamefont{and}
  \bibinfo{author}{\bibfnamefont{J.}~\bibnamefont{Girrbach}},
  \bibinfo{journal}{Phys. Rev.} \textbf{\bibinfo{volume}{D81}},
  \bibinfo{pages}{076001} (\bibinfo{year}{2010}), \eprint{1002.0227}.

\bibitem[{\citenamefont{Gunion et~al.}(1988)\citenamefont{Gunion, Haber, and
  Sher}}]{Gunion:1987qv}
\bibinfo{author}{\bibfnamefont{J.~F.} \bibnamefont{Gunion}},
  \bibinfo{author}{\bibfnamefont{H.~E.} \bibnamefont{Haber}}, \bibnamefont{and}
  \bibinfo{author}{\bibfnamefont{M.}~\bibnamefont{Sher}},
  \bibinfo{journal}{Nucl. Phys.} \textbf{\bibinfo{volume}{B306}},
  \bibinfo{pages}{1} (\bibinfo{year}{1988}).

\bibitem[{\citenamefont{Borzumati et~al.}(1999)\citenamefont{Borzumati, Farrar,
  Polonsky, and Thomas}}]{Borzumati:1999sp}
\bibinfo{author}{\bibfnamefont{F.}~\bibnamefont{Borzumati}},
  \bibinfo{author}{\bibfnamefont{G.~R.} \bibnamefont{Farrar}},
  \bibinfo{author}{\bibfnamefont{N.}~\bibnamefont{Polonsky}}, \bibnamefont{and}
  \bibinfo{author}{\bibfnamefont{S.~D.} \bibnamefont{Thomas}},
  \bibinfo{journal}{Nucl. Phys.} \textbf{\bibinfo{volume}{B555}},
  \bibinfo{pages}{53} (\bibinfo{year}{1999}), \eprint{hep-ph/9902443}.

\bibitem[{\citenamefont{Park}(2010)}]{Park:2010wf}
\bibinfo{author}{\bibfnamefont{J.-h.} \bibnamefont{Park}}
  (\bibinfo{year}{2010}), \eprint{1011.4939}.

\bibitem[{\citenamefont{Noth and Spira}(2010)}]{Noth:2010jy}
\bibinfo{author}{\bibfnamefont{D.}~\bibnamefont{Noth}} \bibnamefont{and}
  \bibinfo{author}{\bibfnamefont{M.}~\bibnamefont{Spira}}
  (\bibinfo{year}{2010}), \eprint{1001.1935}.

\bibitem[{\citenamefont{Guasch et~al.}(2003)\citenamefont{Guasch, Hafliger, and
  Spira}}]{Guasch:2003cv}
\bibinfo{author}{\bibfnamefont{J.}~\bibnamefont{Guasch}},
  \bibinfo{author}{\bibfnamefont{P.}~\bibnamefont{Hafliger}}, \bibnamefont{and}
  \bibinfo{author}{\bibfnamefont{M.}~\bibnamefont{Spira}},
  \bibinfo{journal}{Phys. Rev.} \textbf{\bibinfo{volume}{D68}},
  \bibinfo{pages}{115001} (\bibinfo{year}{2003}), \eprint{hep-ph/0305101}.

\bibitem[{\citenamefont{Crivellin et~al.}(2011)\citenamefont{Crivellin, Hofer,
  and Rosiek}}]{Crivellin:2011}
\bibinfo{author}{\bibfnamefont{A.}~\bibnamefont{Crivellin}},
  \bibinfo{author}{\bibfnamefont{L.}~\bibnamefont{Hofer}}, \bibnamefont{and}
  \bibinfo{author}{\bibfnamefont{J.}~\bibnamefont{Rosiek}}
  (\bibinfo{year}{2011}), \eprint{1103.4272}.

\end{thebibliography}

\end{document}